\documentclass[notitlepage,aps,prb,onecolumn,groupedaddress]{revtex4-1}

\usepackage{graphicx}  
\usepackage[caption=false]{subfig}
\usepackage{dcolumn}   
\usepackage{bm}        
\usepackage{amssymb}   
\usepackage{comment}
\usepackage{hyperref}
\usepackage{color}
\usepackage{amsmath}
\usepackage{mathrsfs}
\usepackage{mathcomp}
\usepackage{textcomp}
\usepackage{dsfont}
\usepackage{esint}
\usepackage{braket}
\usepackage{textgreek}

\DeclareMathOperator{\sgn}{sgn}
\DeclareMathOperator{\diag}{diag}
\newcommand\dd{\mathrm{d}}
\newcommand\ii{\mathrm{i}}
\newcommand\ee{\mathrm{e}}
\makeatletter
\def\ExtendSymbol#1#2#3#4#5{\ext@arrow 0099{\arrowfill@#1#2#3}{#4}{#5}}

\begin{document}

\title{Theory of photoinduced Floquet Weyl semimetal phases}

\author{Xiao-Xiao Zhang$^{1}$}
\author{Tze Tzen Ong$^{1,2}$}
\author{Naoto Nagaosa$^{1,2}$}
\affiliation{$^1$Department of Applied Physics, The University of Tokyo, 7-3-1 Hongo, Bunkyo-ku, Tokyo 113-8656, Japan}
\affiliation{$^2$RIKEN Center for Emergent Matter Science (CEMS), 2-1 Hirosawa, Wako, Saitama 351-0198, Japan}


\makeatother

\begin{abstract}
The Weyl semimetal exhibits various interesting physical phenomena because of the Weyl points, i.e., linear band-crossings. We show by Floquet theory that a linearly polarized light applied to a band insulator can induce controllable Weyl points. In a tight-binding model, we classify different types of photoinduced Weyl points that lead to a rich phase diagram characterized by the Chern number defined on each momentum slices of the bulk states. Taking into account the nonequilibrium electron distribution, we calculate and explain the nonmonotonous anomalous Hall conductivity in terms of the light frequency controlled shift of Weyl points' position, which also allows us to examine the conductivity's dependence on the driving electric field.
\end{abstract}

\maketitle

\section{Introduction}\label{Sec_Intro}
There has been a great surge of interest recently by the community in considering realistic realizations of (linear) band touchings in three dimensions (3D) in solids, which has been a topic of basic importance since the early days of quantum mechanics\cite{vonNeumann1929,Herring1937}. With the broad and varied efforts in topological classification of quantum phases of matter\cite{Topo1,Topo2,Topo3} and Dirac physics\cite{DiracFermion}, originating from the first two-dimensional (2D) material (graphene\cite{graphene}), Weyl fermions\cite{Weyl1929} have emerged in various condensed matter systems, including quantum phase transitions between the topological insulator and band insulator phases in 3D\cite{Weyl2007}, and in the semimetallic electronic structure of pyrochlore iridates\cite{Weyl2011}. This has been followed by a flurry of other theoretical proposals that break time-reversal and/or inversion symmetry\cite{WeylwithT,ReviewQi,ReviewBurkov}, as well as quadratically dispersing double-Weyl fermions with chirality $\pm 2$\cite{QuadraticWeyl1,QuadraticWeyl2,QuadraticWeyl3} together with Coulomb interaction effects therein\cite{QuadraticWeylInteraction1,QuadraticWeylInteraction2,QuadraticWeylInteraction3} and so on. In fact, experimental signatures of Weyl fermions have most recently been successfully observed in experiments on TaAs\cite{seeWeyl1,seeWeyl2} and in a photonic crystal\cite{seeWeyl3}. It is appealing not only as a 3D counterpart of the 2D Dirac physics in graphene, which leads to topologically protected momentum-space monopoles  of the Berry phase\cite{EEMF0,EEMF2}, but also because of other new phenomena as a result of the chiral anomaly\cite{Adler,Bell&Jackiw,Nielson-Ninomiya}. Unlike a topological insulator\cite{TFTTR,TFTTRerr}, such systems can exhibit a spacetime-dependent $\theta$-term, resulting in axion electrodynamics and various concomitant topological responses\cite{thetaWeyl,ReviewQi,ReviewBurkov}, including the anomalous Hall effect (AHE)\cite{AHE1,AHE2} and the chiral magnetic effect (CME)\cite{predictCME,CME1}.

The AHE can be interpreted as the aggregate of the edge mode transport of many 2D quantum Hall planes aligned perpendicular to a third direction in the topologically nontrivial region confined by the Weyl points. However, the CME, which requires some imbalance between separated Weyl points, has two closely related versions, either based on intrinsic energy difference $b_0$ or extrinsic chemical potential difference $\mu_5$ between opposite-chirality Weyl points. While the extrinsic one has been experimentally confirmed by negative magnetoresistance in Dirac/Weyl semimetals\cite{seeCME1Dirac,seeCME2Weyl,seeCME3Dirac}, the intrinsic one, first regarded as an equilibrium current solely under static magnetic field provokes debate\cite{noCME,yesCME} since the total integral of Berry curvature should vanish at equilibrium. Instead, it is later resolved to belong to the more general natural optical activity for noncentrosymmetric materials in the transport limit\cite{Pesin,*Moore1}.

%

Usually topological or quantum phases of matter are achieved via searching for materials with particular features, or through various kinds of band engineering techniques. Among others, the operation of applying light has gained strong attention in recent years, both theoretically and experimentally. Indeed, there has been a long history of exploring to what extent application of light is able to change or exploit material properties. It ranges, for example, from ac transport to the abundant nonlinear phenomena\cite{Boyd}, which are essential to many optical devices in application. Often, such scenarios of light irradiation are challenging because of their time-dependent, nonequilibrium, and nonuniversal nature. Some fundamental questions regarding the fate of equilibration in either integrable or nonintegrable systems still remain open\cite{nonequ1,nonequ2}. Fortunately, in some simpler steady-state situations, much progress has been made towards the characterization of rich quantum phases of matter induced by applying periodically driving electromagnetic fields\cite{Oka0,photoHall,photoMajorana,Kitagawa,FloquetTI,Xing,photoAHE,Aoki,Devereaux}, including quantum Hall effect, Majorana fermions, topological insulators, and so on. Especially, the band structure of the edge modes in a photoinduced topological insulator was experimentally observed\cite{Gedik}. Recently, it is found that nonequilibrium effects can be partly handled with fermion or boson bath to stabilize the desired steady-state\cite{FloquetBath}. In addition, effects of periodic driving on (de)localization in many-body localized quantum states have begun to attract attention\cite{FloquetMBL1,FloquetMBL2}.

Motivated by recent rapid developments, we find it appealing to address the possibility of inducing a Weyl semimetal from a band insulator through the effect of linearly polarized electric field. The merit of this scenario comes from the photon absorption and emission by electrons in solids, which naturally dresses or even extensively modifies the original electronic band structure. In place of the exact quantum electrodynamics description of the polariton generated by the dipole interaction coupling electrons and photons, we prefer the more concise Floquet theory\cite{Floquet1,Floquet2,Floquet3} corresponding to the classical limit of the driving light. 
The photon-dressed band structure comprises only one aspect of the steady-state problem, while the other being the nonequilibrium redistribution of the electron occupations among the newly formed bands, which, however, is crucial to the majority of physical observables. To this end, in addition to providing a positive answer to the possibility, we also study the photoinduced version of the anomalous Hall effect, serving as a connection to realistic and detectable experimental features since many quantum phases of matter manifest themselves via unique transport properties.
This new photoinduced quantum phase, together with its sign-changing and nonmonotonous anomalous Hall conductivity controlled by driving frequency, is interesting in its own right in terms of fundamental science, and is as well its potential for enabling optoelectronic device applications by taking advantage of the fast-switching among different transport signatures.

The paper is organized in the following manner. Sec.~\ref{Sec_protected} discusses the prototypical example of a photoinduced Weyl point formed from a band insulator, which is generated by optically pumping a valence band across the band gap to touch and hybridize with the conduction band. Sec.~\ref{sec: Lattice model} presents a 3D lattice model for realizing such a system, and discusses the high symmetry lines along which the photoinduced Weyl points are located. The Chern number of the different Floquet bands, and the topological charge of the photoinduced Weyl points along the different high symmetry lines are explained in Sec.~\ref{Sec_Chern}, where we also show the phase diagram of the Chern number of the system as a function of the pumping frequency $\Omega$. Finally, we compute the anomalous Hall conductivity using a Floquet-Keldysh formalism in Sec.~\ref{Sec_Hall}, which allows us to take into account the nonequilibrium occupation of the different Floquet bands. The Hall conductivity displays interesting cusp-like points and nonmonotonous behaviour as a function of the pumping frequency $\Omega$, which is due to the creation/annihilation of photoinduced Weyl points and the nonequilibrium occupation of the Floquet bands. We then conclude with a discussion of experimental detection of the photoinduced Weyl points and Hall conductivity.

\section{Photoinduced Weyl points}\label{Sec_Weyl}
\subsection{Topologically protected Weyl point}\label{Sec_protected}
In the Floquet theory, the quantum mechanical problem of a temporally periodic Hamiltonian $H(t)$ with frequency $\Omega$ is transformed to solving the time-independent eigenvalue problem of a new operator $H_F=H(t)-\ii\hbar\partial_t$ in the Floquet-Hilbert space, i.e., a direct product between the ordinary Hilbert space and the space of temporally periodic functions. The eigenenergies, so-called Floquet band structure, are thus periodic in integer multiples of $\hbar\Omega$.
Apart from moving upward or downward the electronic bands via photon absorption or emission, usually the (infinitely many) shifted bands will also get hybridized through the periodically driving term in the Hamiltonian. In most cases, band crossings will be lifted and modified to anticrossings. Here, we show an exception to this general expectation, wherein we perturbatively apply linearly polarized light, $F(t) = V \cos{\Omega t} \,\sigma_x$, to a Weyl point Hamiltonian
\begin{equation}
\label{eq: Weyl Hamiltonian}
H_0 = v_F \vec{k}\cdot\vec{\sigma}.
\end{equation}

The Pauli matrices, $\vec{\sigma}$, here refer to pseudospin orbital degrees of freedom, and the temporally periodic Hamiltonian $H(t)=H_0+F(t)$ is decomposed in the Floquet-Hilbert space in terms of its Fourier components $H^{mn} = \frac{1}{T} \int_0^T {\dd t H(t) \ee^{\ii (m-n)\Omega t}}$ in which $\Omega = \tfrac{2\pi}{T}$ and the superscripts denote the Floquet indices. We readily see that this Hamiltonian becomes block tridiagonal, i.e., only $H^{nn} = H_0$ and $H_\pm \equiv H^{n\mp 1,n}=\frac{V}{2}\sigma_x$ are nonzero. The time-reversal symmetry (TRS) operator for the orbital pseudospins is $\mathcal{T} = K$, where $K$ is the complex conjugation operator. Hence, the Weyl Hamiltonian breaks TRS, as $k_{x(z)} \sigma_{x(z)}$ is odd under $\mathcal{T}$, while the photo-driven hybridization $H_{\pm}$ is time-reversal invariant as we are applying linearly polarized light, which is time-reversal invariant. This is in contrast to the early proposal of photo-driven AHE in graphene via application of time-reversal symmetry breaking circularly polarized light \cite{Oka0}. However, in both cases, the full Floquet Hamiltonian $H^{mn}$ has to break TRS in order to generate a nonzero Hall conductance, i.e. an AHE. 

A similar argument applies to the lattice model that we will consider in Sec.~\ref{sec: Lattice model}. It is well-known that a system has to break either TRS or inversion symmetry in order to have Weyl points. We point out that this model starts from a time-reversal symmetry broken ground state, whose band structure has an intrinsic chirality, which we tap into via optical pumping to create new photoinduced Weyl points. Since we are applying linearly polarized light, we do not consider here the alternate possibility of pumping a noncentrosymmetric system, as TRS has to be broken in order to have a finite AHE.

The harmonic drive $F(t)$ enters through $H_\pm$ in the Floquet Hamiltonian, and Floquet replicas separated by $\pm m$ photon(s) are coupled with a relative strength $\left(\frac{V}{\Omega}\right)^m$; hence, for $\frac{V}{\Omega} \ll 1$, perturbation theory is well defined. Let us first consider the self-energy effects of the optical pumping on the $n$ Floquet band. For $H^{nn}$, the inverse of the bare Green's function reads $g_{0,n}^{-1} = \ii\omega - n\Omega - H^{nn}$, where we have included the photon energy $n \Omega$. To lowest order, the effective Green's function for Floquet replica $n$: $g_{1,n} = \left( g_{0,n}^{-1} - H' \right)^{-1}$, in which the correction to the unperturbed Hamiltonian $H_0$ reads
\begin{equation*}
 H'  = H_- g_{0,n-1} H_+ + H_+ g_{0,n+1} H_- 
   \approx \frac{v_F V^2}{2\Omega^2} \left[
\begin{array}{cc}
  -k_z & k_+ \\
 k_- & k_z \\
\end{array}
\right],
\end{equation*}
wherein we have taken the low-energy limit of $\ii\omega \rightarrow 0,v_F k \ll \Omega$.
Thus, for driving $V$ is sufficiently small, the sole effect of the periodic drive is to renormalize Fermi velocity of the original isotropic Weyl fermion dispersion without opening any gaps nor modifying the chirality of the Weyl point. Since linearly polarized light does not affect the topological charge of the original Weyl points, the other mechanism via which the Chern number of the system can be modified is the creation of new photoinduced Weyl points. 

We are interested in optically controlling the AHE of 3D band insulators and Weyl semimetals, and the total Hall conductivity of the 3D system can be understood as the sum of the quantized anomalous Hall effect for each 2D $k_y$-$k_z$ slice. We assume the band structure of the undriven system is intrinsically chiral, which is modelled using a two band Hamiltonian. In each 2D $k_y$-$k_z$ slice, this can be viewed as a conduction and valence band $\pm E(k_y, k_z)$, separated by a mass gap $2 \tilde{m}(k_x)$. At high symmetry points in each 2D $k_y$-$k_z$ slice, the {\it undriven} helical band structure can be described by a massive Weyl Hamiltonian, $\sum_{(k_y, k_z) \in \mathrm{2D \, BZ}} v_F (k_y \sigma^y + k_z \sigma^z) + \tilde{m} \sigma^x$, which can be derived by Taylor expansion as shown in Sec.~\ref{Sec_classify}. Therefore, the underlying band structure has an intrinsic helicity, which upon photo-irradiation, gives a chiral matrix element, $v^x_{12(21)}$, that hybridizes the optically-pumped $n = -1$ conduction band and $n = 0$ valence band. Specifically, when we apply an electric field along the $\hat{x}$ direction, this will give rise to a $\sigma^x$ perturbation, with $v^x_{12} = A(\cos \theta_0 \cos \phi_0 + i \sin \phi_0)$ and $v^x_{21} = (v^{x}_{12})^{*}$, being the matrix elements of $\sigma^x$ between the original valence and conduction bands. Here, $\cos \theta_0 = \tfrac{v_F k_z}{\sqrt{\tilde{m}^2 + v_F^2(k_y^2 + k_z^2)}}$ and $\tan \phi_0 = \tfrac{v_F ky}{\tilde{m}}$.

\begin{figure}
  \scalebox{0.7}{\includegraphics{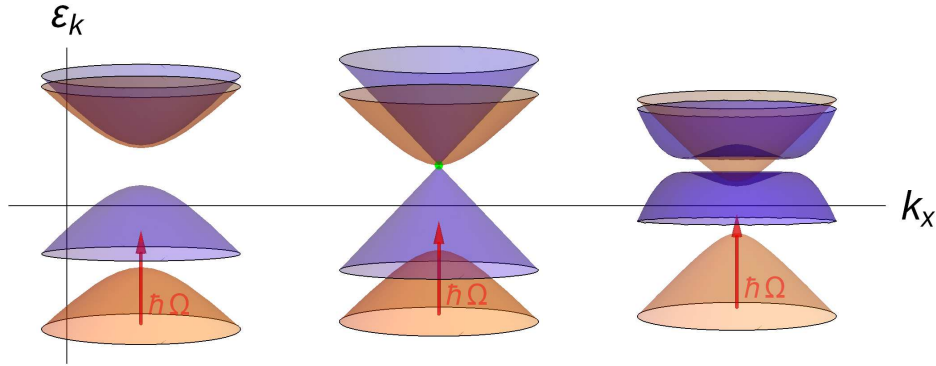}}
\caption{(Color online) Illustration of three typical cases of photoinduced band shifting and hybridization (drawn in the $k_y$-$k_z$ plane) as a function of momentum $k_x$. Without the driving term, we have the gapped orange upper and lower bands. Lower orange band absorbs one photon of frequency $\Omega$ (red arrow) and gets shifted upward to the purple lower band. The two purple bands are the Floquet bands (after hybridization) formed between the foregoing ascended one-photon-dressed band and the original orange upper band. (LEFT) The gap is larger than the light frequency $\Omega$ and hence no band overlap occurs. (MIDDLE) The gap equals to the light frequency at a certain $k_x$ position and the green band-touching gets reshaped linearly and remains gapless under driving by the optical perturbation. (RIGHT) The gap is smaller than the light frequency. Anticrossing gaps open around the ring of band overlap due to the hybridization introduced by the driving term.}\label{fig:bandIllustration}
\end{figure}
The formation of the photoinduced Weyl points between the $n = 0$ Floquet band, with higher index Floquet bands, can also be described in an effective two-band model, and for simplicity, let us consider the 2D Hamiltonian that describes the coupling between the $n = -1$ Floquet conduction band and the $n=0$ Floquet valence band\cite{Morimoto1},
\begin{equation}
\label{eqn:2D Hamiltonian}
H^{(2D)}  =  \begin{pmatrix}
E(k_y, k_z) - \Omega & A v^{x}_{12} \\ 
A v^{x}_{21} & -E(k_y, k_z) \end{pmatrix}  
               =  d_0 \mathbf{1} + \vec{d} \cdot \vec{\sigma} 
\end{equation}
with the Floquet band energy of the driven system given by $\varepsilon^{\pm}(\vec{k}) = d_0(\vec{k}) \pm \sqrt{\delta^2 + |A|^2 |v^x_{12}|^2}$. Here, $\delta$ is the optically-driven overlap between the extrema of the $n=0$ and $n=-1$ Floquet bands, and $|A| \propto \frac{V}{\Omega}$ is the vector potential strength corresponding to the optical pumping field. When the pumping frequency $\Omega$ is smaller than the gap $2 \tilde{m}$, the 2D optically pumped Floquet band structure remains massive as shown in Fig~\ref{fig:bandIllustration} (LEFT). On the other hand, when $\Omega = 2 \tilde{m} + 2 \delta$, one has $d_0 = - \tilde{m} - \delta$, $d_z = -\delta$, $d_x = A \cos \theta_0 \cos \phi_0$ and $d_y = -A \sin \phi_0$. Especially, when the light exactly bridges the gap as shown in Fig.~\ref{fig:bandIllustration} (MIDDLE), i.e., $\delta = 0$, its spectrum is dominated by the linear $|A| |v^x_{12}|$ hybridization term, giving a photoinduced linear dispersion of a prototypical Weyl point (green point in Fig.~\ref{fig:bandIllustration}). When the band overlaps $\delta > 0$, anticrossing massive ring forms as shown in Fig.~\ref{fig:bandIllustration} (RIGHT). Finally, the chirality $\chi$ of the photoinduced Weyl point, i.e. the sign of the topological charge, is determined by the chirality of the undriven $n = 0$ band at the high symmetry lines, which enter through the matrix elements $v^x_{12(21)}$. 

In summary, as illustrated in Fig.~\ref{fig:bandIllustration}, this periodic drive $F(t)$ hybridizes all adjacent photon-dressed bands, thereby affecting the whole band structure, and gaps out all other band crossings away from the high symmetry lines. Apart from the one-photon process illustrated and discussed, photoinduced Weyl points can also form via emission or absorption of multiple virtual photons, i.e. $n$-photon-process. Moreover, we stress that, for any given light frequency, bridging the gapped Hamiltonian to generate a Weyl dispersion is not a fine-tuning phenomenon in a lattice model for real materials simply because the gap $2\tilde{m}(k_x)$ varies continuously as a function of $k_x$ in momentum space. This is in the same spirit that in 3D one can vary three momenta to find generic Weyl points. These points will be made concrete in Sec. \ref{Sec_classify} and the condition for photoinduced formation of a Weyl point is given in Sec.~\ref{Sec_jumpDiagram}.

\subsection{Realization in a lattice model}\label{sec: Lattice model}
In order to realize this system of photoinduced Weyl points, we adopt a 3D two-band model
\begin{equation}\label{Hamiltonian0}
H_{\vec{k}} = [2t_x(\cos{k_x}-\cos{k_0}) + m (\kappa - \cos{k_y} - \cos{k_z})]\sigma_x + 2t_y\sin{k_y}\sigma_y + 2t_z\sin{k_z}\sigma_z,
\end{equation}
wherein we interpret the $\vec{\sigma}$ matrices as psuedospins for orbital degrees of freedom.
Inspecting the energy dispersion relation of this two-band model, we know that for each $k_y$-$k_z$ plane, the $\sigma_y,\sigma_z$ terms in Eq.~\eqref{Hamiltonian0} vanish linearly at the high symmetry points $\tilde{\Gamma} = (0, 0)$, $\tilde{X} = (\pi, 0)\,,(0,\pi)$ and $\tilde{M} = (\pi, \pi)$, and these potential gap-closing points form the foregoing set of high symmetry lines along $\hat{k}_x$. Henceforth for simplicity, we only talk about three high symmetry lines since $(\pi,0)$ and $(0,\pi)$ share the same property. Furthermore, the existence of massless Weyl points requires that the $\sigma_x$ term also vanishes, i.e., $k_x$ satisfying \[2t_x(\cos{k_x}-\cos{k_0}) + m (\kappa+\Lambda) = 0,\] for $\Lambda= - 2, 0, 2$, at $\tilde{\Gamma}$, $\tilde{M}$ and $\tilde{X}$ respectively. For instance, when $\kappa=2$, if $m>t_x$ then there are only two Weyl points at $(\pm k_0,0,0)$\cite{Ran_model} while if $m<t_x$, there are even more. On the other hand, once we set $m$ and $\kappa$ large enough such that $m(\kappa-2)>2t_x (1 + \cos k_0)$, the system becomes a totally gapped band insulator at half filling. 

\begin{figure}[h!]
\subfloat[Weyl point at $\tilde{\Gamma}$ point when $k_x = -1.32$]{
\label{fig:typeA} 
\begin{minipage}[c]{0.40\textwidth}
\centering
  \scalebox{0.45}{\includegraphics{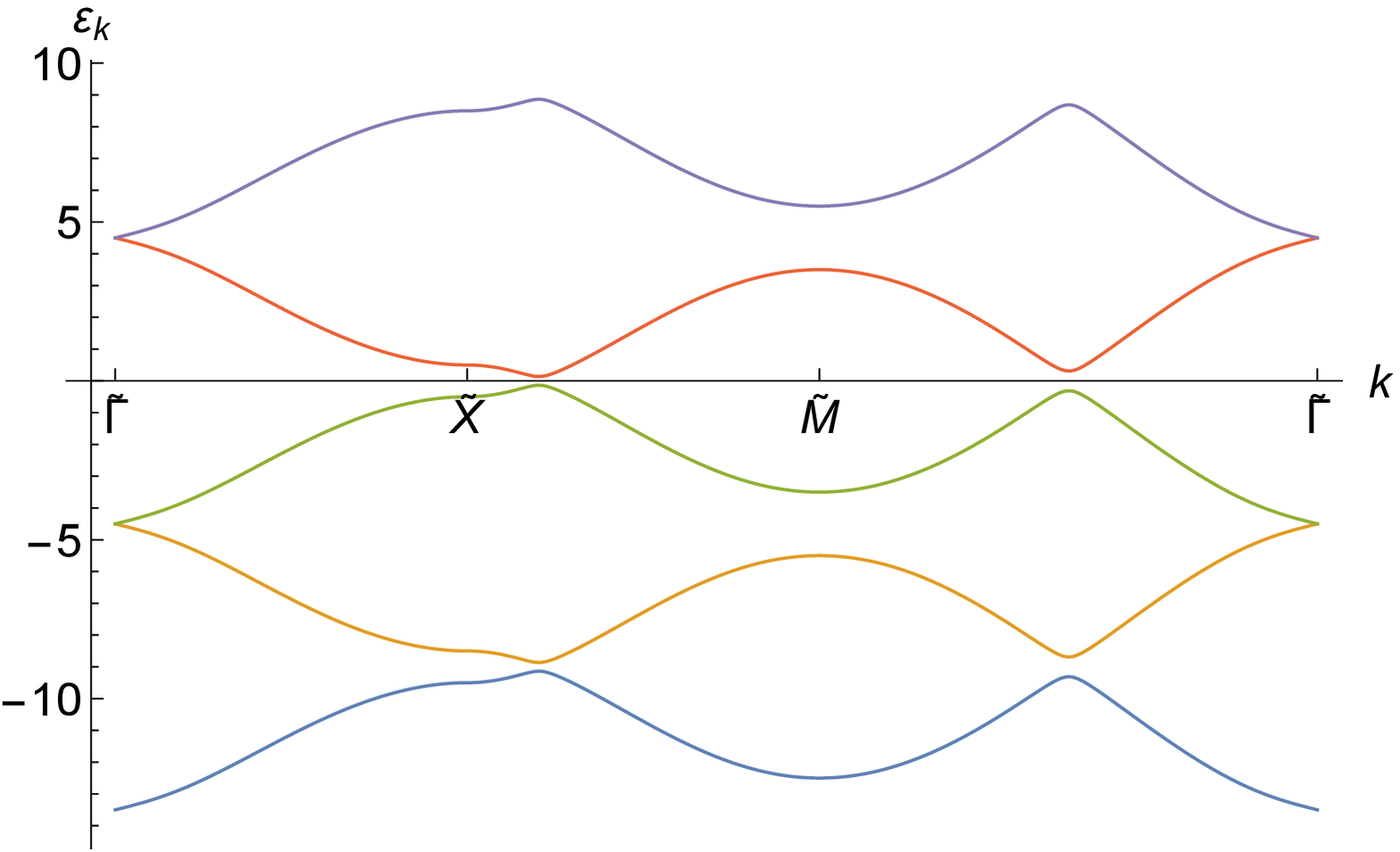}}
\end{minipage}
  \hfill
\begin{minipage}[c]{0.28\textwidth}
\centering
  \scalebox{0.45}{\includegraphics{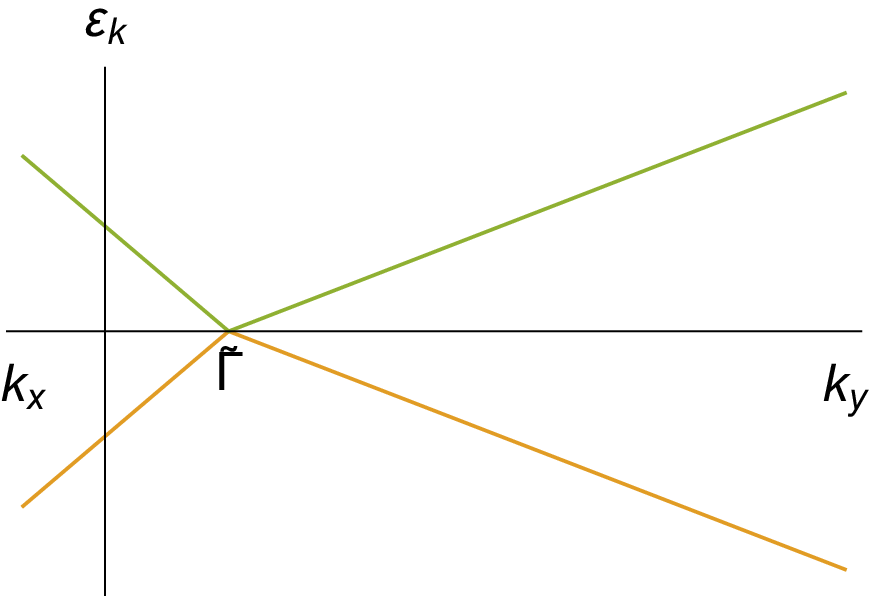}}
\end{minipage}
  \hfill
\begin{minipage}[c]{0.32\textwidth}
\centering
  \scalebox{0.45}{\includegraphics{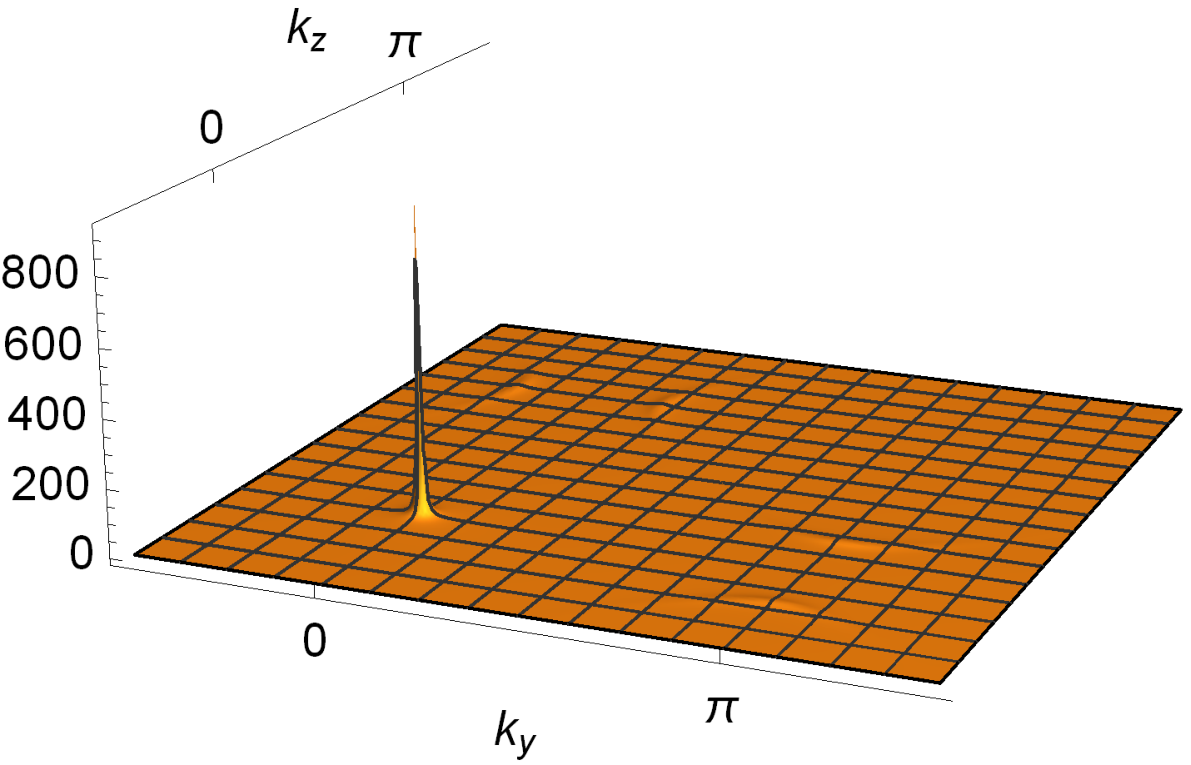}}
\end{minipage}}\\[-10pt]
\subfloat[Weyl point at $\tilde{M}$ point when $k_x = -0.72$]{
\label{fig:typeB} 
\begin{minipage}[c]{0.40\textwidth}
\centering
  \scalebox{0.45}{\includegraphics{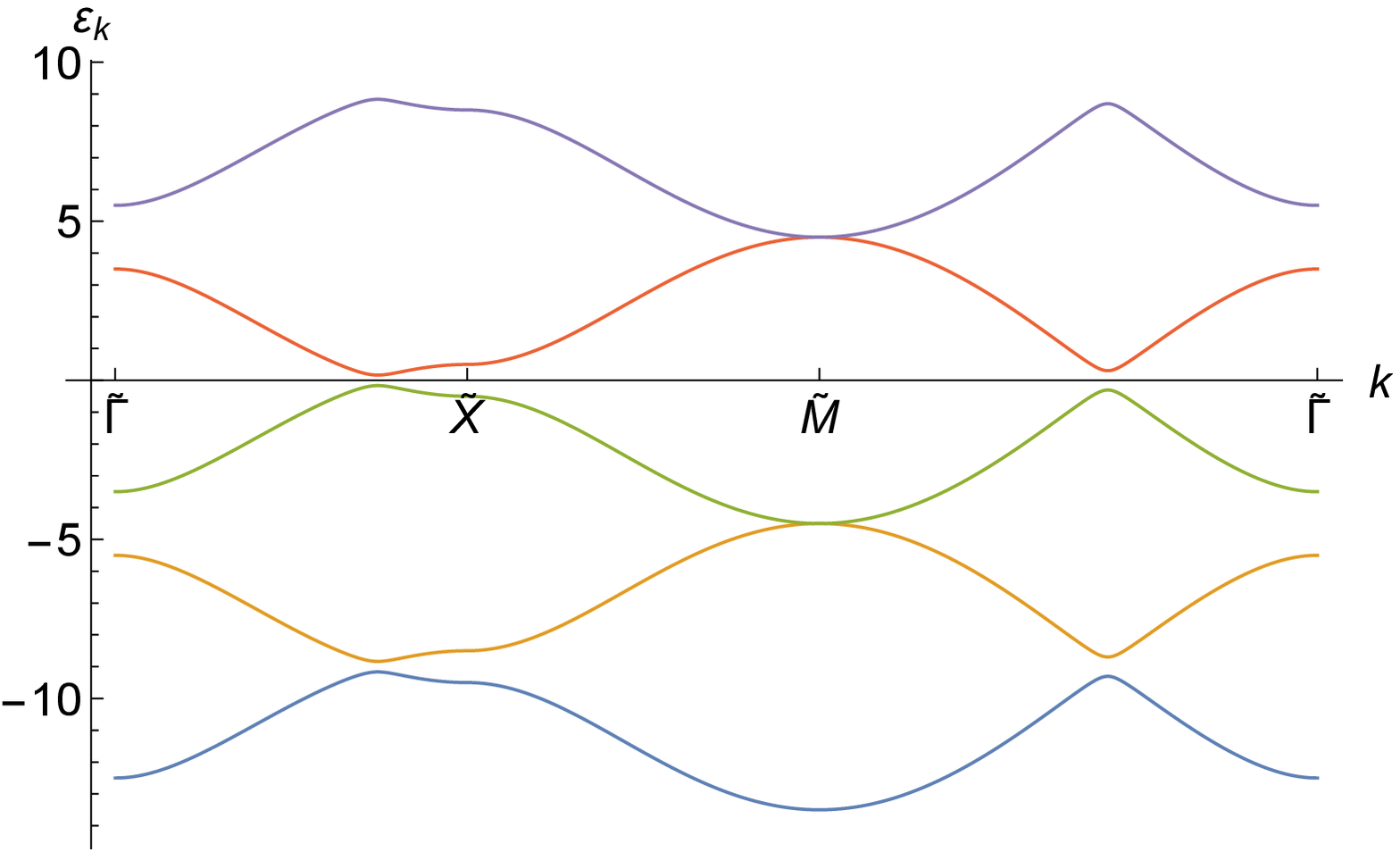}}
\end{minipage}
  \hfill
\begin{minipage}[c]{0.28\textwidth}
\centering
  \scalebox{0.45}{\includegraphics{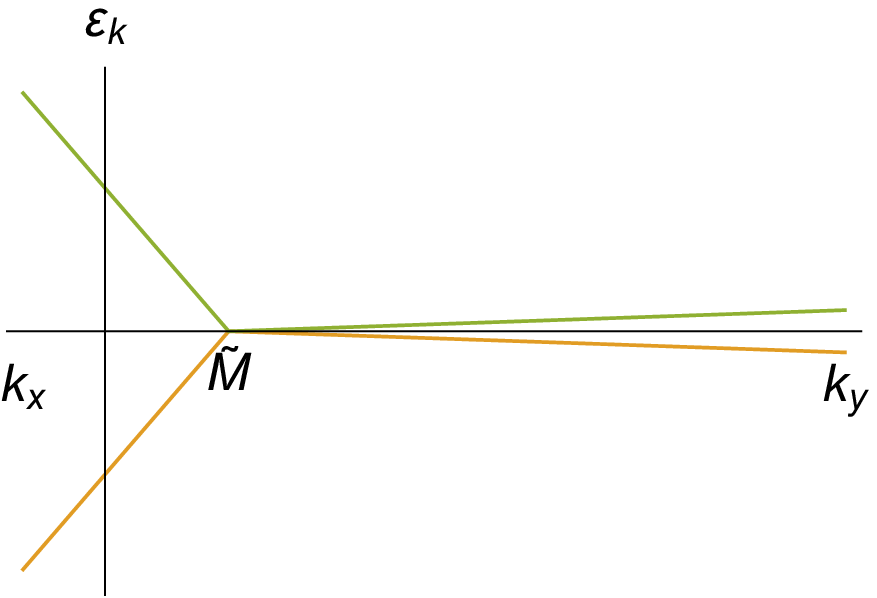}}
\end{minipage}
  \hfill
\begin{minipage}[c]{0.32\textwidth}
\centering
  \scalebox{0.45}{\includegraphics{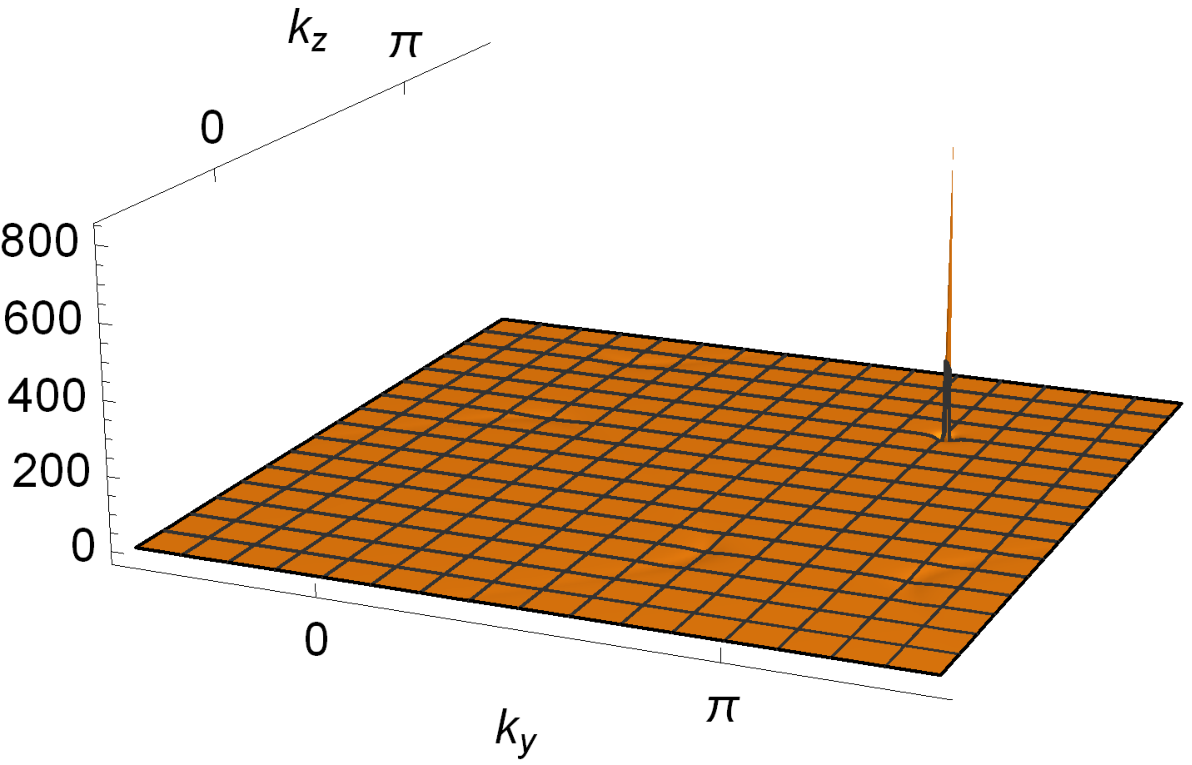}}
\end{minipage}}\\[-10pt]
\subfloat[Weyl point at $\tilde{X}$ point when $k_x = -1.05$]{
\label{fig:typeC} 
\begin{minipage}[c]{0.40\textwidth}
\centering
  \scalebox{0.45}{\includegraphics{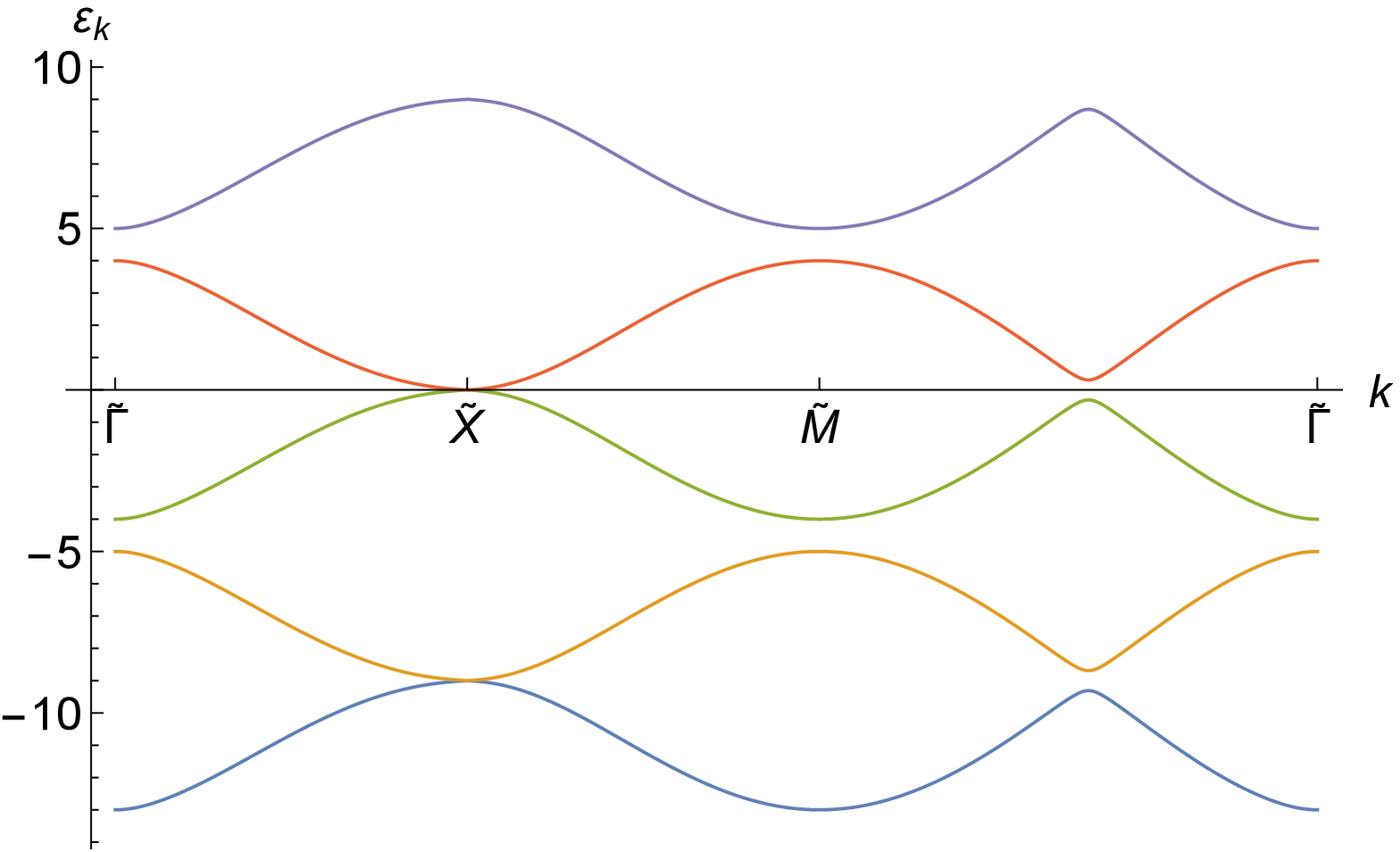}}
\end{minipage}
  \hfill
\begin{minipage}[c]{0.28\textwidth}
\centering
  \scalebox{0.45}{\includegraphics{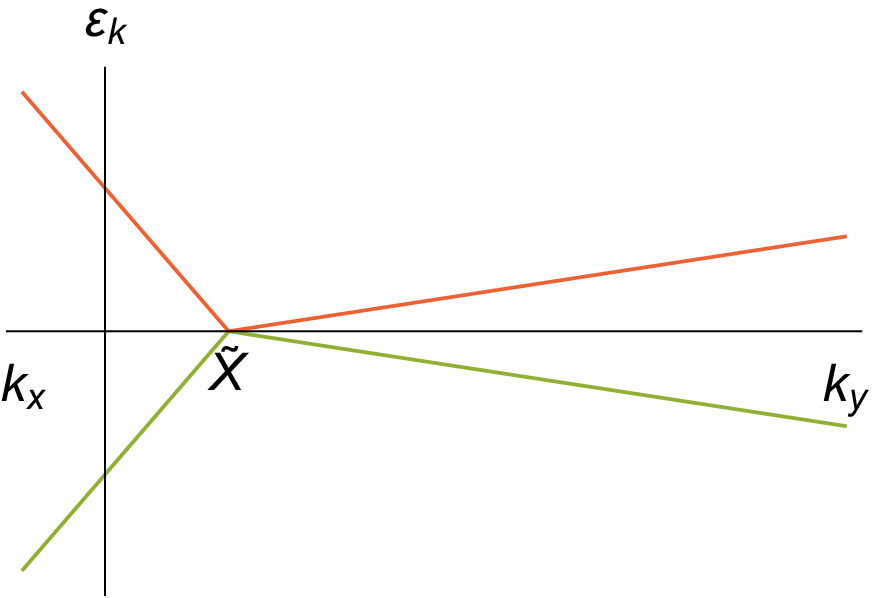}}
\end{minipage}
  \hfill
\begin{minipage}[c]{0.32\textwidth}
\centering
  \scalebox{0.45}{\includegraphics{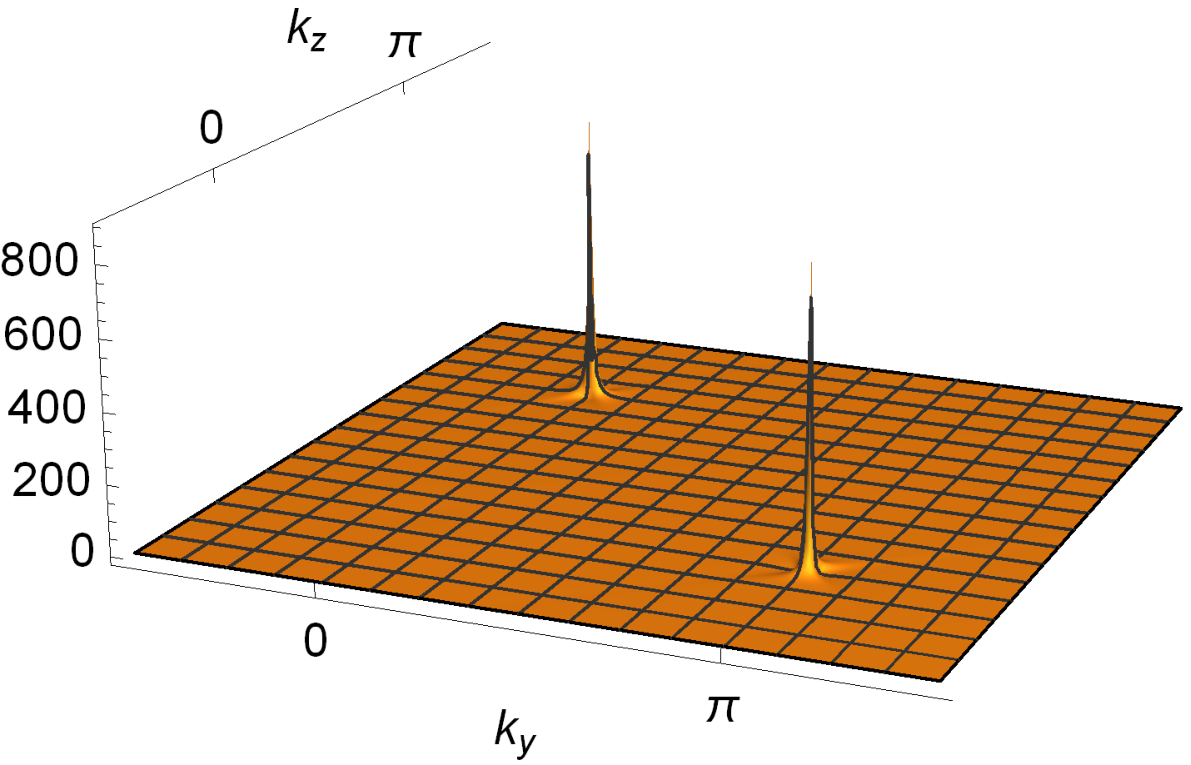}}
\end{minipage}}\\[-0.5pt]
\subfloat[2D projected 1st BZ in the $k_y$-$k_z$ plane.]{\label{fig:BZ}
\centering
  \scalebox{0.45}{\includegraphics{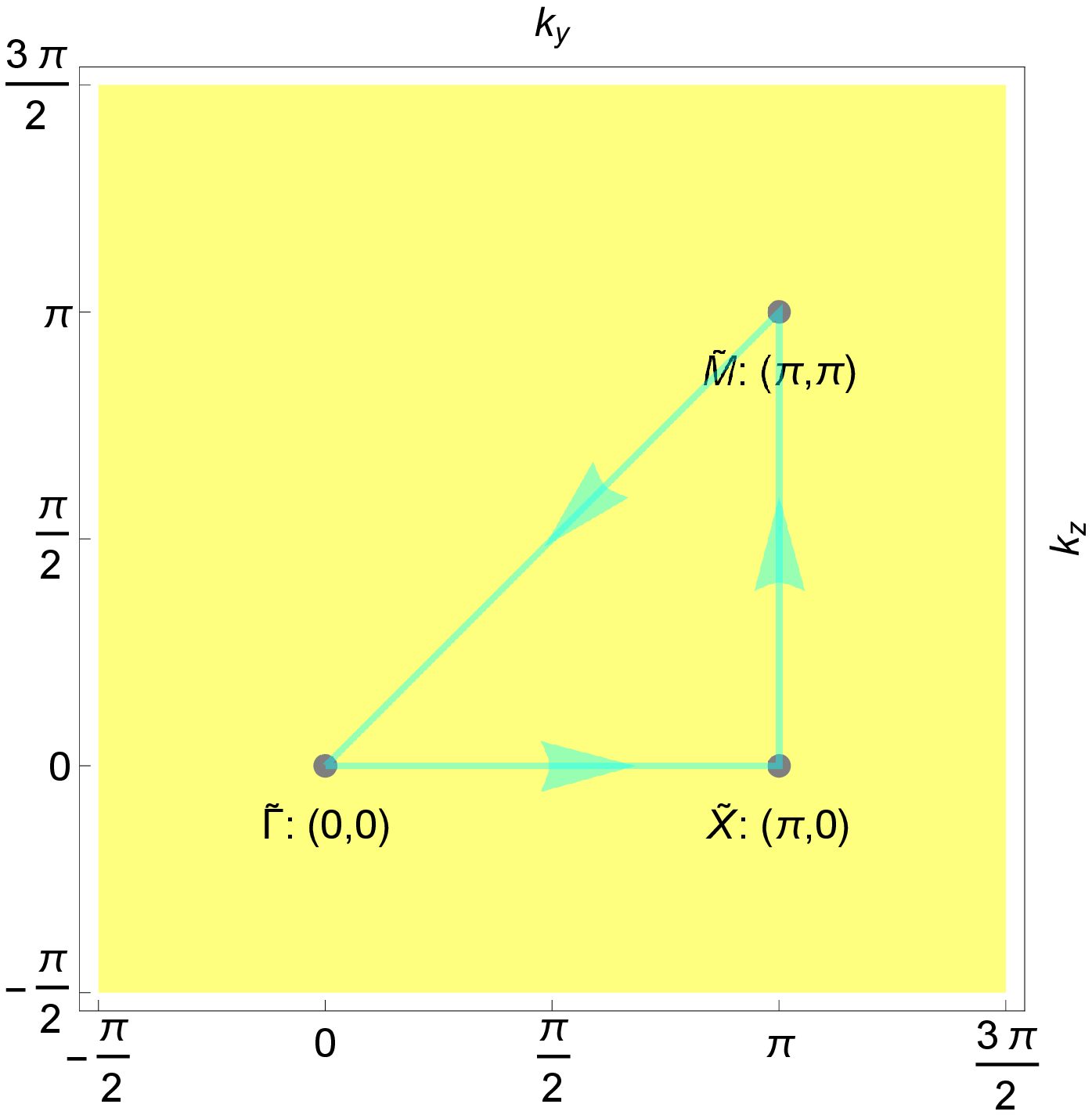}}
  }
  \caption{(a)-(c) $\Omega=9$ periodic drive applied to the band insulator. LEFT: Five adjacent Floquet bands near the zero energy along special lines in $k_y$-$k_z$ plane depicted by the cyan triangle in (d). Middle: Linear dispersion along $k_x$ (towards the left) and $k_y$ (towards the right) axes close to the Weyl point. RIGHT: Berry curvature of the green $(0,-)$ band over the first Brillouin zone in $k_y$-$k_z$ plane when $k_x$ lies slightly above the Weyl point.}\label{fig:Fband}
\end{figure}

The effective mass at the three different high symmetry points, $\tilde{m}(\tilde{\Gamma})$, $\tilde{m}(\tilde{M})$ and $\tilde{m}(\tilde{X})$, for a particular $k_y$-$k_z$ slice, is given by the coefficient of the $\sigma_x$ term, and it is clear that $\tilde{m}(\tilde{M}) > \tilde{m}(\tilde{X}) > \tilde{m}(\tilde{\Gamma})$. The band structure at these high symmetry points is then described by the 2D Hamiltonian Eq.~\eqref{eqn:2D Hamiltonian}, and the formation of photoinduced Weyl points at these points follows the discussion after Eq.~\eqref{eqn:2D Hamiltonian}. For simplicity, we henceforth set $t_x=t_y=t_z=1,k_0=\frac{\pi}{2}$ and in general consider $m=2,\kappa=4$ (insulator) in our calculation unless otherwise stated. In Sec.~\ref{Sec_Chern} we will mention the case of a Weyl semimetal, for the parameters $m=1.2,\kappa=2$, with Weyl points at $(\pm \frac{\pi}{2}, 0, 0)$.
We then turn on the periodic drive $F(t)$ in this system, and we now have many Floquet replicas of Eq.~\eqref{Hamiltonian0}, which are coupled by off-diagonal blocks $H_\pm$. This Hamiltonian in the Floquet-Hilbert space can be numerically solved by exact diagonalization with truncation at some Floquet index. We found that this tight-binding model indeed realizes the scenario of photoinduced Weyl points, and we show the results for $\Omega = 9$ in Fig.~\ref{fig:Fband}, where we fix the value of $k_x$ and show the band structure along the high symmetry lines in the first $k_y$-$k_z$ Brillouin zone (1st BZ) (Fig.~\ref{fig:BZ}). 

For the above-mentioned parameters for the band insulator, the effective masses $\tilde{m}(\tilde{\Gamma}) = 6$, $\tilde{m}(\tilde{X}) = 10$ and $\tilde{m}(\tilde{M}) = 14$ for the $k_x = 0$ plane. Hence, for a pumping frequency of $\Omega = 9$, photoinduced Weyl points will occur between the $n = 0$ and $n =1$ Floquet bands at $\tilde{\Gamma}$, between the $n = 0$ and $n =2$ Floquet bands at $\tilde{X}$, and between the $n = 0$ and $n =3$ Floquet bands at $\tilde{M}$, at different $k_x$ planes. Specifically, these three cases occur at $k_x = -1.32$, $k_x = -0.72$ and $k_x = -1.05$ respectively, as shown in Fig.~\ref{fig:Fband}, for $2\tilde{m}(\tilde{\Gamma}) = \Omega$, $2\tilde{m}(\tilde{X}) = 2\Omega$, and $2\tilde{m}(\tilde{M}) = 3\Omega$ respectively. The band touchings remain gapless with linear dispersion; the Fermi velocity along $k_y$ and $k_z$ are equal as $y,z$ are symmetric in the model, but it is anisotropic with respect to $k_x$. These linear dispersions are shown in the middle column of Fig.~\ref{fig:Fband}. Apart from this, there are only anticrossings due to hybridization by optical pumping, and the gaps opened at these more ubiquitous anticrossings are always of the order $V \left(\frac{V}{\Omega}\right)^{n}$, where $n$ is the difference of Floquet indices between the two anticrossed bands. A similar hierarchy of such dynamic gap-openings also occurs in the 2D Dirac fermion system\cite{FloquetHierarchy}.

A Weyl point is also a source or sink, i.e., a magnetic monopole, of Berry phase flux in the momentum space. According to the Kubo formula, the Berry curvature or the curl of the Berry connection $\vec{a}_\alpha$ of Floquet band $\alpha$ reads
\begin{equation}\label{eq:BerryCurvature}
(\nabla\times \vec{a}_\alpha)_{x}(\vec{k}) = \;\sum_{\beta\neq \alpha} {\ii \; \frac{\braket{\alpha\vec{k}|\partial_{k_y}h_0|\beta\vec{k}} \braket{\beta\vec{k}|\partial_{k_z}h_0|\alpha\vec{k}} - \textrm{H.C.} }{(\varepsilon_{\alpha\vec{k}} - \varepsilon_{\beta\vec{k}})^2}},
\end{equation}
We further plot the Berry curvature distribution in the $k_y$-$k_z$ BZ, in the right part of Fig.~\ref{fig:Fband}. We see that the Berry curvature is strongly concentrated, essentially delta-function-like, around the special points $\tilde{\Gamma},\tilde{M},\tilde{X}$ as it should be for a momentum-space monopole. The topological charge for a monopole is defined by the total flux of the lower $\alpha^{\mathrm th}$-band passing through a closed surface $\vec{S}$ that encloses the monopole, $Q = \tfrac{1}{4 \pi} \int d \vec{S} \cdot (\nabla_{\vec{k}} \times \vec{a}_{\alpha}(\vec{k}))$. We carried out the calculation in the 1st BZ $[-\pi,\pi]_{k_x}\times [-\frac{\pi}{2},\frac{3\pi}{2}]_{k_{y}/ k_z}^2$, rendering the monopoles away from the edges of the $k_y$-$k_z$ 1st BZ. Lastly, according to our calculation, each photoinduced Weyl point exists in company with another Weyl point, located symmetrically with respect to $k_x = 0$ on the high symmetry lines along $\hat{k}_x$. Moreover, as seen from the corresponding Berry curvature plot when $k_x>0$, they form a monopole-antimonopole pair in momentum space, hence satisfying the Nielson-Ninomiya theorem\cite{Nielson-Ninomiya} as in the case of a conventional Weyl semimetal. This results from the inversion symmetry with respect to $k_x$ in the model, which is not broken by the driving term. All the above features hold not only in the case we presented, but also in all the other cases with different parameters of the model, which manifestly shows the generality of the proposed phenomenon.

\section{Chern number diagram}\label{Sec_Chern}
The Chern number provides important information of topological classification and properties of the system, such as in the case of quantum Hall effect, topological insulator\cite{Topo1,Topo2}, and also topological superconducting systems. From the viewpoint of treating the simple cubic 1st BZ of the 3D bulk system as stacking up many slices of 2D square $k_y$-$k_z$ 1st BZ at continuous $k_x$ points, one can define 'slice Chern numbers' at every $k_x$,
\begin{equation}
\label{eqn: slice Chern number}
Ch(k_x) = \frac{1}{2 \pi} \int \dd k_y \dd k_z (\nabla \times \vec{a}_{\alpha})_{x} (\vec{k})。 
\end{equation}
Such consideration can be naturally extended to Floquet bands as well. Sweeping through a gap-closing Weyl point or a monopole in the momentum space will change the Chern number regardless of whether it is a conventional Weyl semimetal or a photoinduced Weyl system. Similarly, in our driven system, the Chern number jump $\Delta Ch$ for a particular Floquet band upon increasing $k_x$ is simply determined by the topological charge of the monopole and its position in momentum space. 
The topological charge $\chi$ of the monopole in a two-band model is defined by the total flux of the Berry curvature of the lower photoinduced helical band. Therefore, a particular band has a corresponding Chern number jump, $\Delta Ch=\mp \chi$, if it is the upper/lower photoinduced helical band of the monopole. For instance, the green band is the upper (lower) band for the $\tilde{\Gamma}$-point ($\tilde{X}$-point) monopole in Fig.~\ref{fig:typeA} (Fig.~\ref{fig:typeC}).
 
We assign index $n=0$ to the Floquet replica nearest to zero energy, e.g., the green and red bands in Fig.~\ref{fig:Fband}. Then we use $n>0$ ($n<0$) indices to denote the replicas above (below) the $n=0$ replica. To distinguish the upper/lower band within the $n$th replica, we will denote it as $(n,\pm)$ band. For an intra-replica Weyl point formed between the $(n,\pm)$ bands, the $(n, \pm)$ bands will naturally refer to the upper and lower photoinduced helical bands respectively. However, an inter-replica Weyl point may also form between bands from adjacent replicas, e.g., between $n$ and $n - 1$, and in such a case, the $(n,-)$ band will be the upper helical band, while the $(n-1,+)$ band will be the lower helical band for that particular Weyl point. For example, Fig.~\ref{fig:typeA} shows an inter-replica Weyl point at $\tilde{\Gamma}$, and Fig.~\ref{fig:typeC} shows an intra-replica Weyl point at $\tilde{X}$.

We integrate the Berry curvature over the $k_y$-$k_z$ 1st BZ and henceforth focus on the Chern number and Weyl points within $k_x\in [-\pi,0]$ of the $(0,-)$ band (green band in Fig.~\ref{fig:Fband}), which could form monopoles with either the $(0,+)$ band (red band in Fig.~\ref{fig:Fband}) or the $(-1,+)$ band (orange band in Fig.~\ref{fig:Fband}). These two cases will then have a corresponding $\Delta Ch = \chi$ and $\Delta Ch = -\chi$, respectively. In addition, because of the aforementioned symmetric distribution in the region $k_x\in[-\pi,\pi]$, we will have corresponding opposite-chirality monopoles for $k_x \in [0, \pi]$.

\subsection{Classification of the photoinduced Weyl points}\label{Sec_classify}
\begin{figure}
\subfloat[]{
\label{fig:BZinfo} 
\begin{minipage}[c]{1.0\textwidth}
\centering
  \scalebox{0.45}{\includegraphics{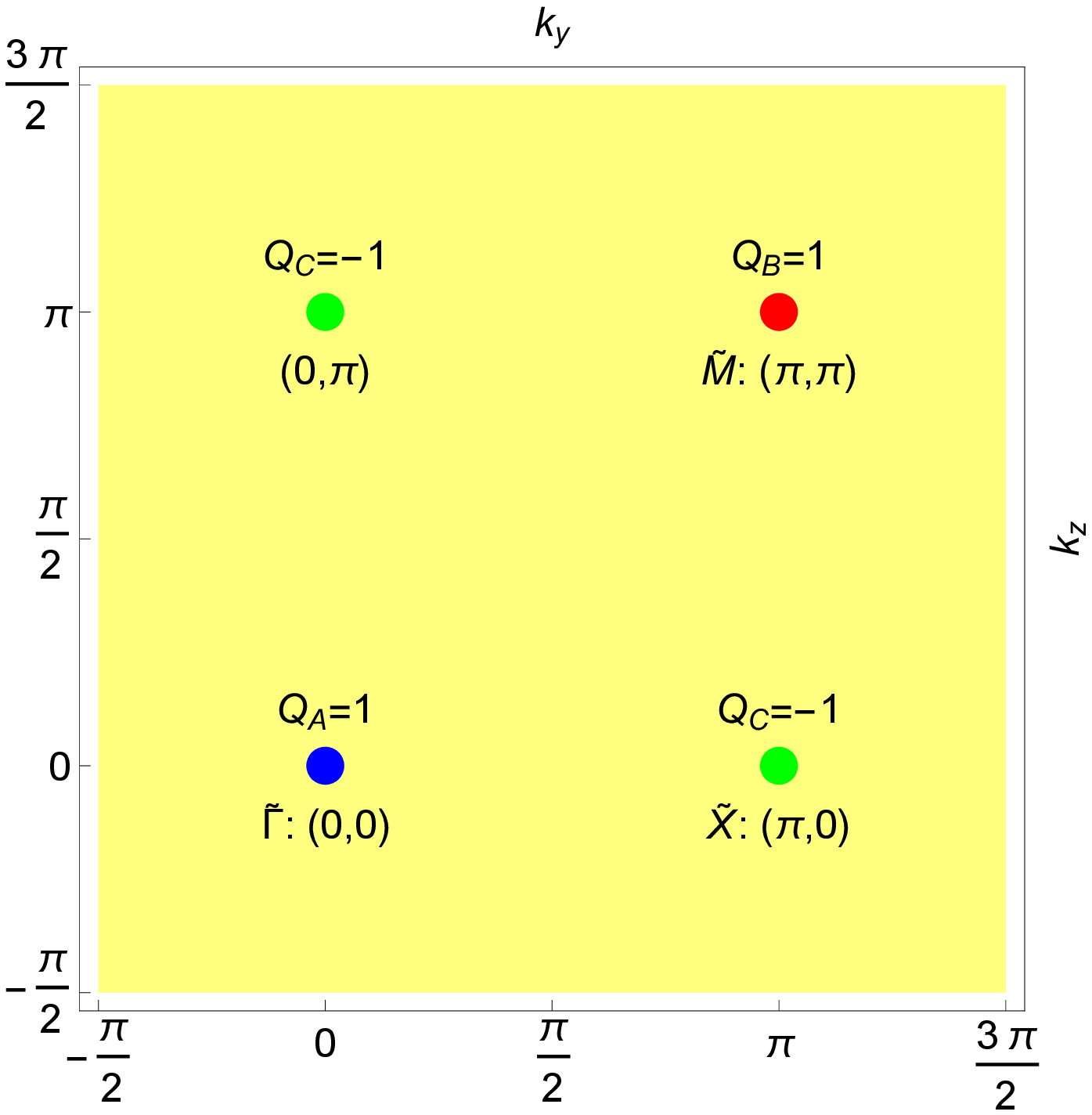}}s
\end{minipage}}\\[-0pt]
\subfloat[]{
\label{fig:lineDiagram8Original} 
\begin{minipage}[c]{1.0\textwidth}
\centering
  \scalebox{0.45}{\includegraphics{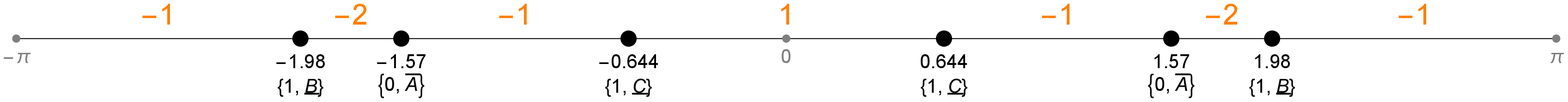}}
\end{minipage}}\\[-1pt]
  \caption{(Color online) (a) Classification of the photoinduced Weyl points in the 2D 1st BZ in the $k_y$-$k_z$ plane. Four high symmetry lines along the $k_x$ direction in the full 3D 1st BZ are signified by colored dots, where photoinduced Weyl points (monopoles in momentum space) are possible to reside. Type-$A$ or type-$B$ monopole has topological charge $+1$. Type-$C$ monopole has topological charge $-1$ and appears always in pair as shown by the two green dots since $k_y$-$k_z$ are on the same footing in the model. (b) Typical example of Chern number and monopoles along $k_x$ axis. $\Omega=8$ for the model initially possessing two Weyl points at $\vec{k}=(\pm\frac{\pi}{2},0,0)$.}
\end{figure}

The chirality $\chi$ of a Weyl point or the charge $Q$ of a Berry phase monopole is given by $\chi = \sgn [\det(v_{ij})]$ for a two-band model $h(\vec{k}) = \sum_{ij} {v_{ij}k_i\sigma_j}$\cite{Volovik}. Since the optical perturbation only renormalizes $v_F$, as described in Sec.~\ref{Sec_protected}, the chirality of the $n = 0$ Floquet band and of the matrix elements $v^x_{12(21)}$ remains invariant; hence, the chirality $\chi$ can be calculated from the \textit{undriven} Hamiltonian \eqref{Hamiltonian0}.
We expand it at a generic point $\vec{K}^0 = (k_x^0,k_y^0,k_z^0)$ and get 
\begin{equation*}\label{Hamiltonian1}
\begin{split}
H_{\vec{k}} =& [2t_x(\cos{k_x^0}-\sin{k_x^0}\Delta k_x-\cos{k_0}) + m (\kappa - \cos{k_y^0} - \cos{k_z^0} + \sin{k_y^0}\Delta k_y + \sin{k_z^0}\Delta k_z)]\sigma_x \\
&+ 2t_y(\sin{k_y^0}+\cos{k_y^0}\Delta k_y)\sigma_y + 2t_z(\sin{k_z^0}+\cos{k_z^0}\Delta k_z)\sigma_z + O({\Delta k}^2).
\end{split}
\end{equation*}
For any point $\vec{K}^0$ on the aforementioned high-symmetry lines, we have 
\begin{equation}\label{Hamiltonian2}
H_{\vec{k}} = f(\Delta k_x,\vec{K}^0)\sigma_x + 2t_y\cos{k_y^0}\Delta k_y \sigma_y + 2t_z\cos{k_z^0}\Delta k_z \sigma_z,
\end{equation}
 where we define $f(\Delta k_x,\vec{K}^0) = 2t_x(\cos{k_x^0}-\sin{k_x^0}\Delta k_x-\cos{k_0}) + m (\kappa - \cos{k_y^0} - \cos{k_z^0})$ and $f(\Delta k_x = 0,\vec{K}^0)$ is nothing but the $\tilde{m}$ discussed throughout Sec.~\ref{Sec_Weyl}. 
The band gap, given by $\Delta = 2\tilde{m}$, is illustrated in Fig.~\ref{fig:bandIllustration} by the energy difference between the orange bands' extrema. After photon-dressing, the gap is eliminated and turned into a common extremum indicated by the green point. Therefore, in terms of the Hamiltonian \eqref{Hamiltonian2}, we have $v_{ij} = \diag(-2t_x\sin{k_x^0},2t_y\cos{k_y^0},2t_z\cos{k_z^0})$ around such a band-touching point $\vec{K}^0$ without taking hybridization into account. One readily obtains $\chi = \sgn(-8t_xt_yt_z\sin{k_x^0}\cos{k_y^0}\cos{k_z^0})$, which leads to the following classification as shown in Fig.~\ref{fig:BZinfo},
\begin{itemize}
\item[$A$]: one $\chi=1$ monopole at the $\tilde{\Gamma}$-point $(k_y^0,k_z^0)=(0,0)$
\item[$B$]: one $\chi=1$ monopole at the $\tilde{M}$-point $(k_y^0,k_z^0)=(\pi,\pi)$
\item[$C$]: two $\chi=-1$ antimonopoles at the two inequivalent $\tilde{X}$-points $(k_y^0,k_z^0)=(0,\pi),(\pi,0)$.
\end{itemize}

The formation of $\chi = -1$ antimonopoles at $\tilde{X}$ is easily understood by a change of sign of the matrix elements $v^x_{12(21)}$ and a change of chirality at $\tilde{X}$ as compared to $\tilde{\Gamma}$. Specifically, from Sec.~\ref{Sec_protected}, $d_y$ changes sign, $d_y \rightarrow A \sin\phi$, thereby changing the rotation direction of the $\vec{d}$ vector and the charge of the photoinduced monopole.
The monopole charges of the $A$, $B$ and $C$ Weyl points given above are defined in terms of the Berry curvature of their lower helical bands as usual. As aforementioned in the beginning of Sec.~\ref{Sec_Chern}, in order to calculate the change in Chern number of a Floquet band, it is necessary to keep track of whether the monopole is over or underneath that band, i.e., intra-replica or inter-replica. Therefore, focusing on the $(0,-)$ band, we mark the intra-replica monopoles with an overline ($\overline{A}$, $\overline{B}$, $\overline{C}$), and the inter-replica monopoles with an underline ($\underline{A}$, $\underline{B}$, $\underline{C}$), signifying whether a Weyl point is between the $(0,-)$ band and $(0,+)$ band, or it is between the $(0,-)$ band and $(-1,+)$ band, which we will refer to as 'over- and under-type' henceforth.

We illustrate this binary nature of the photoinduced Weyl points for a band insulator subjected to an optical pumping with $\Omega = 9$ in Fig.~\ref{fig:Fband}, where Fig.~\ref{fig:typeA} \ref{fig:typeB} \ref{fig:typeC} correspond to type $\underline{A}$, $\underline{B}$ and $\overline{C}$ Weyl points. We provide another example for an original Weyl semimetal at a smaller pumping frequency $\Omega = 8$ in Fig.~\ref{fig:lineDiagram8Original}, which shows the pre-existing and photoinduced Weyl points at the same time. The two type-$\overline{A}$ points are the original Weyl points in the undriven Weyl semimetal, which remains gapless after switching on the light as we expected in Sec. \ref{Sec_protected}. For this particular choice of parameters, the photoinduced Weyl points at $\tilde{M}$ and $\tilde{X}$ both form at the bottom of the $(0,-)$ band; hence they are denoted as $\underline{B}$ and $\underline{C}$ Weyl points at $k_x = \pm 1.98$ and $k_x = \pm 0.644$ respectively. The numerically calculated values of `slice Chern number' $Ch(k_x)$ are displayed as orange integer numbers above the $k_x$ axis, and the Weyl points responsible for the change of $Ch(k_x)$ are displayed as black dots, with their $k_x$ positions marked beneath. The number in curly brackets is the difference of Floquet indices between the two touching bands, and since $\overline{A}$ are the original Weyl points, they are naturally formed by crossings of the original $n = 0$ Floquet bands, while the photoinduced Weyl points are formed between $(0,-)$ and $(-1,+)$, with a difference in Floquet index of 1. Thus, our analysis of chirality together with over- and under-type information fully explains the numerically obtained Chern number jumps.

\subsection{Chern number jump and phase diagram as a function of optical frequency}
\label{Sec_jumpDiagram}
According to the physical picture described in Secs. \ref{Sec_protected} and \ref{Sec_classify}, all the positions of photoinduced Weyl points can be predicted by listing the possible band-touching degeneracies of relevant Floquet bands, which is nothing but solving [provided $E(\vec{k}) \geq 0$]
\begin{equation}\label{eq:listing}
-E(\vec{k})+m\Omega = E(\vec{k})+n\Omega \Rightarrow E(\vec{k})=\frac{m-n}{2}\Omega,
\end{equation} 
where $m,n\in\mathds{Z}$ and $E(\vec{k})$ is the band energy of the undriven Hamiltonian \eqref{Hamiltonian0}  in the same spirit as Sec. \ref{Sec_classify}. Any solution $(k_x,\Omega)$ signifies a photoinduced Weyl point in the driven Floquet system. This implies that in order to generate any new Weyl point, one must have $\Omega < 2E_\textrm{max}$, in which the band energy extremum $E_\textrm{max}=14$ of the original Hamiltonian Eq.~\eqref{Hamiltonian0} for our parameter choice. In addition to the tripartite category $A,B,C$, one more important point is how one can distinguish between the over- and under-type category defined in Sec.~\ref{Sec_classify}. Noticing that any over- (under-) type Weyl point always exists at (below) zero energy, we readily know that this is directly determined by the parity of $m-n$ (referred as 'Floquet difference' henceforth) in Eq.~\eqref{eq:listing}. Thus, we arrive at our final formula of Chern number jump in terms several monopoles (if there are more than one simultaneously) indexed by $i$
\begin{equation}\label{eq:jump_formula}
\Delta Ch = \sum_i{(-)^{m_i-n_i}\chi_i}.
\end{equation}

\begin{figure}
\begin{center}
  \scalebox{0.44}{\includegraphics{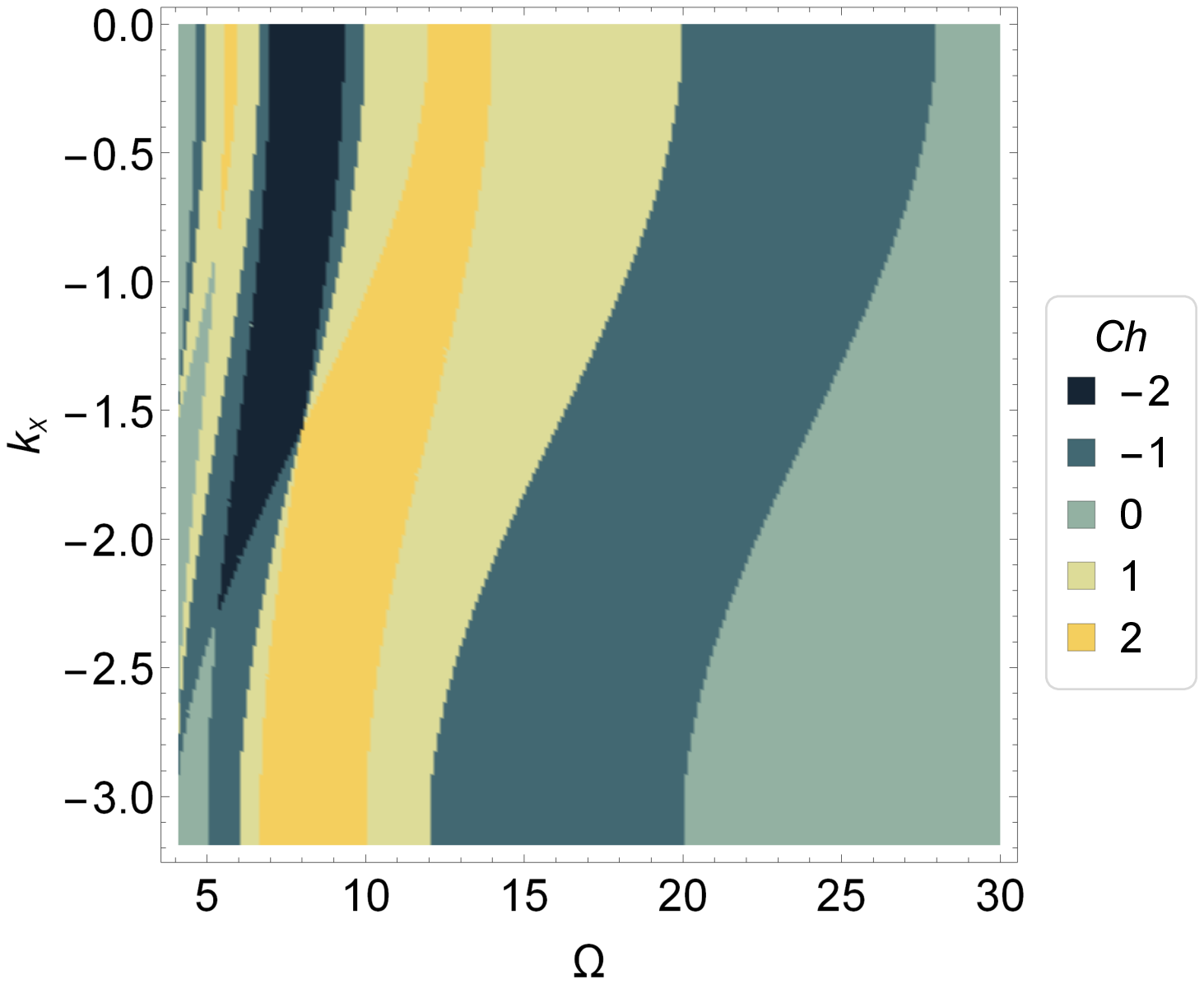}}
  \scalebox{0.317}{\includegraphics{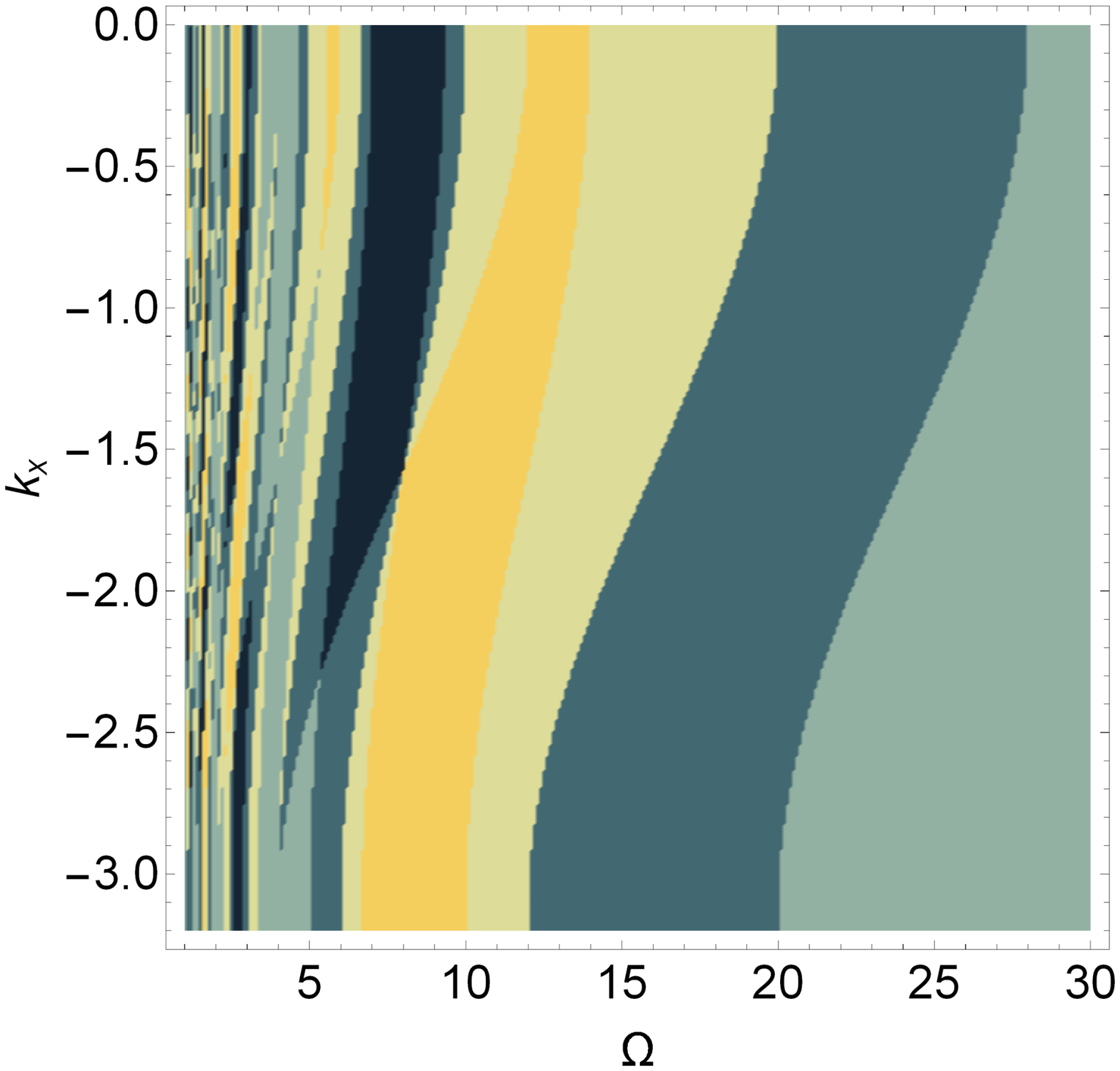}}
  \scalebox{0.317}{\includegraphics{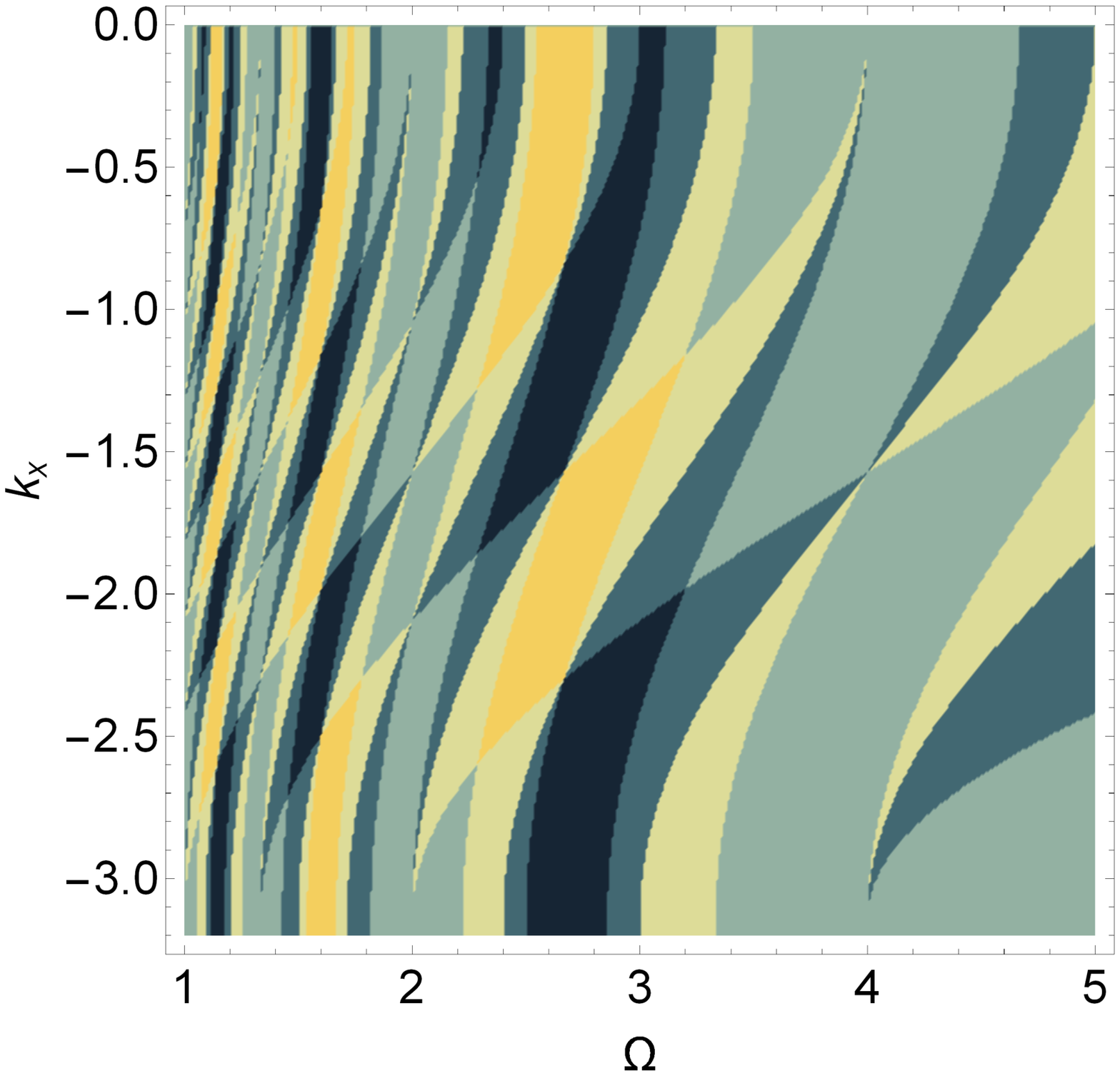}}
  \caption{(Color online) Chern number diagram as a function of $k_x\in[-\pi,0]$ and driving frequency $\Omega$ for the $(0,-)$ band. LEFT: Numerical result via solving the Floquet Hamiltonian for $\Omega\in[4,30]$. Middle: Analytic result for a larger range $\Omega\in[1,30]$. RIGHT: Analytic result for the small frequency region $\Omega\in[1,5]$.}\label{fig:ChernDiagram}
\end{center}
\end{figure}
For instance, it turns out that Eq.~\eqref{eq:jump_formula} can fully reproduce the Chern number variation in Fig.~\ref{fig:lineDiagram8Original} once the Floquet difference information is calculated via solving Eq.~\eqref{eq:listing}, which corresponds to three simple cases
\begin{align*}
&\overline{A}:-E+0\Omega=E+0\Omega \Rightarrow E=0\\
&\underline{B}:-E=E+\Omega \Rightarrow E=\frac{-1}{2}\Omega\\
&\underline{C}:-E=E-\Omega \Rightarrow E(\vec{k})=\frac{1}{2}\Omega.
\end{align*}
In Fig.~\ref{fig:ChernDiagram}, we first show in the left panel our numerical calculation result for the Chern number distribution in the $k_x$-$\Omega$ plane, which is obtained by solving the Floquet Hamiltonian and integrating over the Berry curvature. Because of the aforementioned inversion symmetry of $k_x$, it suffices to show the diagram for $k_x\in[-\pi,0]$. From Eq.~\eqref{eq:listing}, we know that there always exist many high-photon-number (high Floquet difference) Floquet band touchings for small enough $\Omega$, the number of which diverges when $\Omega \rightarrow 0$. Because the integration requires too long time to converge for the complex Berry curvature distribution, it is formidable to enter the small frequency regime numerically. 

On the other hand, for frequency $\Omega > 2E_\textrm{max}$, all the Floquet replicas are well separated in energy, i.e., no gap-closing nor overlap occurs at all to alter the topological number. That is why when $\Omega > 28$ we observe the same completely zero Chern number as the undriven band insulator model. Therefore, instead of solving the band structure, we are also able to start from this 'separated' situation and use Eq.~\eqref{eq:listing} and Eq.~\eqref{eq:jump_formula} to analytically account for all the photoinduced Weyl points all the way down to the small frequency regime as shown in the rest of Fig.~\ref{fig:ChernDiagram}. It turns out that analytic analysis fully reproduces the numerical result, and indeed, the stripes become denser for smaller frequencies. Also, in terms of the analytic relation Eq.~\eqref{eq:listing}, every stripe line in a Chern number diagram depicts the shift of a Weyl point as we increase the driving frequency. Furthermore, it should only start at $k_x=-\pi$ and end at $k_x=0$, which in fact signifies the creation and annihilation of a Weyl point. Any stripe line that does not exactly reach the edges or is disconnected is a numerical artifact.

We find that certain points in Fig.~\ref{fig:ChernDiagram} are intersections of several stripe lines. This just corresponds to the summation in Eq.~\eqref{eq:jump_formula}, meaning simultaneous Weyl points. Actually, the full relation between slice Chern number jump, $\Delta Ch$, and Weyl point category is as follows
\begin{itemize}
\item $+ 1$: $\overline{A}$ or $\overline{B}$ \,, $- 1$: $\underline{A}$ or $\underline{B}$
\item $- 2$: $\overline{C}$ \,, $+ 2$: $\underline{C}$
\item $+ 4$: $\overline{A}+\overline{B}+\underline{C}$ \,, $- 4$: $\underline{A}+\underline{B}+\overline{C}$.
\end{itemize}
where $\Delta Ch$ takes possible values of $\pm 1\,, \pm 2\,,  \pm 4$. For instance, the three stripe lines intersect at $k_x= -\frac{\pi}{2},\Omega=8$, correspond to the $+4$ case. We have three comments derived from Eq.~\eqref{eq:listing}. 1) For $\Delta Ch = \pm 2$, the naive guess of $\overline{A}+\overline{B}$ ($\underline{A}+\underline{B}$) without $C$ is actually forbidden since same over- and under-type of $A$ and $B$ dictates a $\overline{C}$ or $\underline{C}$ at the same time, giving rise to $\Delta Ch = 0,\pm 4$. 2) Also note that $\Delta Ch = \pm 3$ is impossible since satisfying $\overline{A}+\underline{C}$ or $\overline{B}+\underline{C}$ ($\underline{A}+\overline{C}$ or $\underline{B}+\overline{C}$) automatically implies $\overline{A}+\overline{B}+\underline{C}$ ($\underline{A}+\underline{B}+\overline{C}$). 3) Any other combination simply results in no change of Chern number because of the cancellation between monopoles of opposite chirality. Indeed, in Fig.~\ref{fig:ChernDiagram}, all these claims are not only confirmed by numerically solving the model, but also by the diagrams based on analytic relations, and we see the slice Chern number ranges between $-2$ and $+2$.

\section{Anomalous Hall conductivity}\label{Sec_Hall}
The Chern number diagram encodes rich information of the photo-driven system, especially the topological properties. However, the photoinduced Weyl points  are topological transition points in the Floquet-Hilbert space, instead of the ordinary Hilbert space, which is reflected in the periodicity of the Floquet band structure, as shown in Fig.~\ref{fig:Fband}, with an optically-pumped nonequilibrium electron occupation. For an ordinary Weyl semimetal, one characteristic signature is the AHE, which is essentially understood by considering the cumulative effect of many 2D quantized anomalous Hall layers with nonzero slice Chern numbers. If symmetry allows for a nonzero net Chern number, this should be experimentally measurable. In order to relate our study to more realistic and detectable quantities, we study the AHE of our driven system with photoinduced Weyl points in this section, which is given by the following 3D integral of momentum and summation over Floquet bands
\begin{equation}\label{eq:Hall}
\sigma_{yz} = \frac{e^2}{2\pi h}\int{ \dd^3 \vec{k}\sum_\alpha{\tilde{f}_\alpha(\varepsilon_{\alpha\vec{k}}) (\nabla\times \vec{a}_\alpha)_{x}(\vec{k}) } }.
\end{equation}

This follows from a nonequilibrium generalization of the 2D Hall effect\cite{Oka0}. As the Berry curvature is for the photo-driven Floquet bands, the occupation function $\tilde{f}$ should reflect a nonequilibrium distribution, which, however, is not universal and depends on the driving process in general. Even for a stationary case, the occupation is by no means similar to a conventional Fermi distribution; due to the nonequilibrium irradiation effect, higher Floquet bands will also contain electrons that contribute to conductivity. To be systematic, we combine the Keldysh Green's function techniques\cite{Keldysh1,Mahan} with the Floquet theory and arrive at a Floquet-Keldysh formalism\cite{Oka0,Morimoto1}, in which the Dyson equation reads
\begin{align*}
\begin{pmatrix}
G^R & G^K \\
0 & G^A
\end{pmatrix}^{-1}_{mn}
&=
\begin{pmatrix}
(\omega-n\Omega)\delta_{mn}-H_{mn} & 0 \\
0 & (\omega-n\Omega)\delta_{mn}-H_{mn}
\end{pmatrix}
+\Sigma_{mn}
\end{align*}
 and $\Sigma$ is the self-energy. The lesser Green's function is
$G^<=G^R \Sigma^< G^A$ for stationary states\cite{Keldysh1} and lesser self-energy $\Sigma^< =(\Sigma^R+\Sigma^K-\Sigma^A)/2$. Here, we consider a simple but possibly realistic situation in which the whole system is coupled to a heat reservoir of conventional Fermi distribution at a certain temperature $T$ with a damping factor $\Gamma$
\begin{align*}
\Sigma_{mn}&=
\ii\Gamma \delta_{mn}
\begin{pmatrix}
\frac 1 2 & -1+2f(\omega-m\Omega) \\
0 & -\frac 1 2
\end{pmatrix}.
\end{align*}
The occupation can be given by $\tilde{f}_\alpha(\omega) = \braket{\psi_\alpha^F|\Sigma^<|\psi_\alpha^F}/\ii\Gamma$, wherein the Floquet state $\ket{\psi_\alpha^F}$ is solved from diagonalizing the Hamiltonian $H_{mn}+\delta_{mn}m\Omega$ as we have done previously. We assume the chemical potential of the heat reservoir to be at the middle point, i.e., the zero energy, in the spectrum of the original two-band model.

\subsection{Nonmonotonous Hall conductivity}\label{Sec_conductivity}
For a conventional undriven system at equilibrium, the conductivity at zero temperature would be simply determined by the total Chern number of the bands below the Fermi energy. Furthermore, the Berry curvatures of all adjacent bands cancel out, except for the $(0,-)$ band. This follows from the 'zero sum rule' that all the band Chern numbers add up to zero, which can be proved by a homotopy argument\cite{BerryConservation}. 
However, this is not the case for our nonequilibrium system. The Hall conductivity results not only from the over-type Weyl points, but also from the under-type ones, due to the nonequilibrium occupation of all the bands. Thus, all bands matter, requiring us to carry out a numerical calculation, which includes high Floquet index bands, for the Hall conductivity. 

\begin{figure}
\begin{center}
  \scalebox{0.7}{\includegraphics{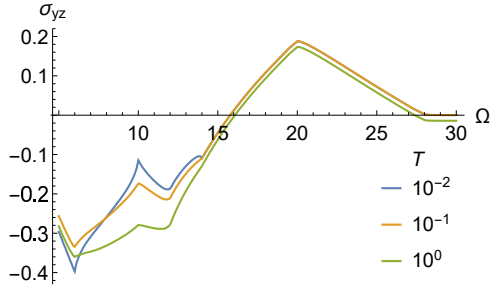}}
  \caption{(Color online) Photoinduced anomalous Hall conductivity v.s. driving frequency $\Omega$. Temperature $T$ is measured in units of hopping energy $t_x$ in the Hamiltonian.}\label{fig:Hall}
\end{center}
\end{figure}

We employ an adaptive algorithm\cite{Genz1,Genz2} to carry out the numerical integration in the 3D 1st BZ. The result shown in Fig.~\ref{fig:Hall} is calculated when fixing the ratio $\frac{V}{\Omega}$ to a relatively small value $\frac{1}{3}$. Since the strength $V$ of the driving term $F(t)$ is proportional to the magnitude of the ac electric field and $|\vec{E}(t)| \propto |\dot{\vec{A}}(t)|$, this means the conductivity variation is plotted at the same strength of vector potential. Similar to the calculation in Sec. \ref{Sec_jumpDiagram}, the small $\Omega$ region becomes numerically inaccessible. In such a system, tuning the driving frequency can continuously deform the distance between any pair of photoinduced Weyl points. This, together with the nonequilibrium occupation, leads in general to non-quantization of the anomalous Hall conductivity as the Weyl points are not fixed at special $k_x$ points. However, we find several noteworthy features such as sign-change, nonmonotonous behaviour and existence of cusps, implying a rich potential for photo-based control of the system in different possible applications.

\begin{figure}
\centering
\subfloat[$\Omega_0=6.0$]{
\label{fig:Omega6} 
\begin{minipage}[c]{1.0\textwidth}
  \scalebox{0.45}{\includegraphics{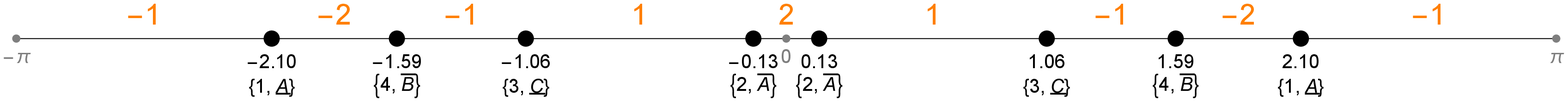}}
  \scalebox{0.45}{\includegraphics{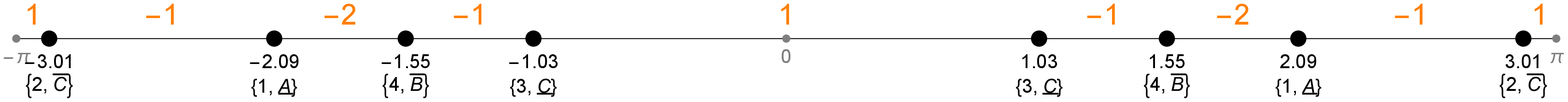}}
\end{minipage}}\hfill
\subfloat[$\Omega_0=10.0$]{
\label{fig:Omega10} 
\begin{minipage}[c]{1.0\textwidth}
  \scalebox{0.45}{\includegraphics{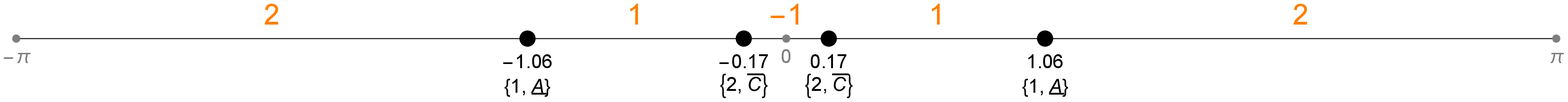}}
  \scalebox{0.45}{\includegraphics{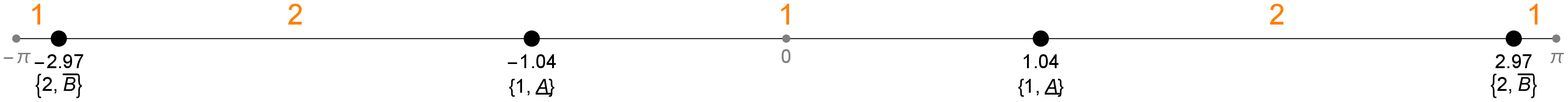}}
\end{minipage}}\hfill
\subfloat[$\Omega_0=20.0$]{
\label{fig:Omega20} 
\begin{minipage}[c]{1.0\textwidth}
  \scalebox{0.45}{\includegraphics{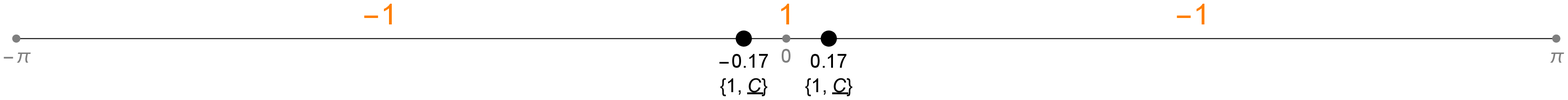}}
  \scalebox{0.45}{\includegraphics{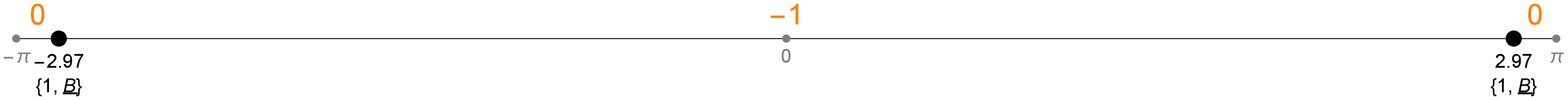}}
\end{minipage}}
  \caption{(Color online) Typical creation and annihilation of photoinduced Weyl points are shown by the Chern number diagrams along $k_x$ direction. Upper (lower) diagram in each subfigure is plotted when the driving frequency $\Omega$ is $3\text{\textperthousand}$ below (above) the frequency $\Omega_0$ at which pair creation and/or annihilation occur. Meanings of various markers are the same as those introduced in Fig.~\ref{fig:lineDiagram8Original} and Sec. \ref{Sec_classify}.}\label{fig:Hall_monopole}
\end{figure}

Although the anomalous Hall conductivity depends on the complete band structure and the corresponding occupation, we are able to understand the key qualitative features. The formation of cusps is due to the creation/annihilation of the Weyl points, shown in the phase diagram Fig.~\ref{fig:ChernDiagram} of the $(0,-)$ band, while the nonmonotonous variation is due to the nonequilibrium occupation of the photoinduced bands. Because of periodicity in energy, the $(0,-)$ band contains all the representative information for the Weyl points.
We first point out the necessary condition for forming a cusp in $\sigma_{yz}(\Omega)$, which is creation at $k_x=\pm\pi$ and/or annihilation at $k_x=0$ of a pair of Weyl points of opposite chiralities. For instance, this can mean annihilation of a pair of $A/ B$ Weyl points at $k_x = 0$ with creation of a pair of $C$ Weyl points at $k_x = \pm \pi$, or annihilation of a pair of $C$ Weyl points at $k_x = 0$ with creation of a pair of $A/ B$ Weyl points at $k_x = \pm \pi$, and these two cases are shown in Fig.~\ref{fig:Omega6} and \ref{fig:Omega10} respectively. This is because the $\Omega$-dependence of the conductivity for type $A/B$ versus $C$ are opposite, as we will explain next using the one-photon process as an example. We point out that both over-type $\overline{A}, \overline{B}, \overline{C}$ and under-type $\underline{A}, \underline{B}, \underline{C}$ Weyl points are formed via the same physical mechanism, and hence, contribute in the same manner to the AHE. The over- and under-type notation refers to the position of the Weyl point with respect to a particular band, which is only necessary for tracking the Chern number phase diagram of that band, e.g. the $(0,-)$ band as shown in Fig.~\ref{fig:ChernDiagram}.

Previewing the results from Sec.~\ref{Sec_Vdependence}, we show that for a type-$A$ Weyl point with upper and lower photoinduced helical bands formed from an $n = -1$ conduction band and and $n = 0$ valence band respectively, the occupation for the upper photoinduced helical band increases with frequency $\Omega$, and decreases for the lower helical band. Hence, increasing $\Omega$ will pump electrons from the lower positive Chern band to the upper negative one for each 2D $k_y$-$k_z$ slice, causing a net decrease in the conductivity. Similar physics underlies the type-$B$ and $C$ Weyl points as well. Specifically, the band structure is inverted (hole-like) at $\tilde{M}$ for type-$B$ Weyl points, and the upper/lower photoinduced helical band are formed from an $n = 0$ valence band/$n = 1$ conduction band respectively, i.e., reversed as compared to type-$A$. Hence, electrons are pumped from the upper photoinduced positive Chern band to the lower negative one upon increasing $\Omega$. Finally, for the type-$C$ Weyl points, the photoinduced upper and lower helical bands are formed from the $n = 0$ valence and $n = -1$ conduction bands. This, together with a negative monopole charge, is reversed compared to type-$A$ Weyl points. 
Thus, increasing $\Omega$ will pump electrons from the upper negative Chern band to the lower positive one, causing an increase in the conductivity. Hence, the optical pumping effect will lead to a decrease of AHE conductivity for type-$A$ and $B$ Weyl points, while increasing it for type-$C$ Weyl points, upon increasing $\Omega$. In case of ambiguity, we point out that the upper and lower photoinduced helical bands used here correspond to the left figure in Fig.~\ref{fig:bandIllustration} while the right one shows the band inversion after crossing a Weyl point. 

Let us briefly illustrate these two points by examining the photo-driven shift of the photoinduced Weyl points in Fig.~\ref{fig:Hall_monopole}. In Fig.~\ref{fig:Omega6}, a pair of type-$A$ Weyl points move towards and annihilate with each other at $k_x=0$ and afterwards a pair of type-$C$ Weyl points are created at $k_x=\pm\pi$ and then move towards $k_x=0$. As explained above, these two discontinuous processes are responsible for the valley-like cusp around $\Omega=6$. Increasing $\Omega$ from less than $6$ will pump more electrons from the lower positive Chern band to the upper negative one for the $A$ Weyl points at $k_x = \pm 0.13$, thereby reducing $\sigma_{yz}(\Omega)$. For $\Omega > 6$, similar effect on the newly created $C$ Weyl points will drive an increase in $\sigma_{yz}(\Omega)$.
Similarly, Fig.~\ref{fig:Omega10} shows a pair of type-$C$ Weyl points' annihilation, followed by creation of a pair of type-$B$ Weyl points, which first increases $\sigma_{yz}(\Omega)$ and then decreases $\sigma_{yz}(\Omega)$ when traversing through $\Omega=10$. Finally, similar reasoning applies to the peak-like cusp around $\Omega=20$ for Fig.~\ref{fig:Omega20}.

In addition, there is a weaker cusp point at $\Omega = 14$ where  annihilation of $\overline{B}$ Weyl points at $k_x = 0$ occurs without a concomitant creation of other Weyl points at $k_x = \pm \pi$. Furthermore, Weyl points of large Floquet index difference have correspondingly smaller contributions as they are higher order in perturbation. 
The complicated band dependence of the occupation number can also moderate the singular contribution from creation or annihilation of Weyl points. Hence the cusp-like effect could be compensated or diminished by the shift of other Weyl points.
Finally, we point out that the type-$B$ Weyl points in Fig.~\ref{fig:Omega20} are the only surviving photoinduced two Weyl points when $\Omega>20$, as $\Omega$ now exceeds the gap at $\tilde{X}$. They are destined to annihilate with each other when $\Omega=28$ as seen in Fig.~\ref{fig:lineDiagram8Original} and mentioned in Sec.~\ref{Sec_jumpDiagram}. Consequently, the anomalous Hall conductivity returns to zero (up to a residual conductivity caused by nonzero temperature) when $\Omega>28$ since the photo-driven system becomes trivial.

Thus, the nonmonotonous profile of this AHE can be accounted for by two simple observations. 1) As the driving frequency $\Omega$ increases, nonequilibrium pumping effects on a Weyl point of type $C$ ($A$ or $B$) leads to an increase (decrease) in $\sigma_{yz}(\Omega)$, regardless of its over- and under-type. 2) The necessary condition for a cusp in $\sigma_{yz}(\Omega)$ is the creation at $k_x=\pm\pi$ and/or annihilation at $k_x=0$ of a pair of Weyl points of opposite chiralities.

\subsection{Conductivity dependence on driving strength $V$}\label{Sec_Vdependence}
In terms of nonlinear optical responses, photo-current $j$ in an inversion-symmetric system under light irradiation ordinarily behaves as\cite{Boyd} $j\propto E^3$ where $E$ is the strength of the ac electric field and proportional to $V$ in our model. Or, more formally, the third-order nonlinear current reads 
\[
j_i=\chi^3_{ij} E(\Omega)E(-\Omega)E_j(\omega=0).
\]
It means that the photoinduced Hall conductivity we calculated should behave as $\sigma\propto E^2$ since it comes from the second-order effect of the ac electric field, and it can be detected when another dc electric field $E_j(\omega=0)$ is applied. From the analysis below, we show that $\sigma \propto E^2$ when $E$ is small compared to the damping rate, $\Gamma$, in the system; whereas, in the regime of large electric field $E \gg \Gamma$, it crosses over to $\sigma \propto E$.

Since the anomalous Hall conductivity of the 3D system is an integral of the Hall conductivity of all the 2D $k_y$-$k_z$ slices, hence, by focusing on a single 2D $k_y$-$k_z$ slice, we are able to elucidate the $\Gamma$ and $E$ dependencies of the Hall conductivity in a similar analysis for the model \eqref{eqn:2D Hamiltonian} \cite{Morimoto1}. The 2D Hamiltonian that describes the anticrossing between the photon-dressed Floquet band of index $n = -1$ conduction band and the $n=0$ valence band one was shown in Eq.~\eqref{eqn:2D Hamiltonian}, and the wave functions for this two-band Floquet model are
$
u_{+}(\vec{k}) = (\cos \frac{\theta}{2} \ee^{-i \phi}, \sin \frac{\theta}{2} )^\mathrm{T}\,,
u_{-}(\vec{k}) = ( -\sin \frac{\theta}{2} e^{-i \phi} , \cos \frac{\theta}{2} )
^\mathrm{T}$,
where $\cos \theta = \frac{d_z}{|\vec{d}|}$ and $\tan \phi = \frac{d_y}{d_x}$. From Eq.~(\ref{eq:Hall}), the Hall conductivity due to the photoinduced Weyl points is then given by the Berry curvature of the two bands, with a corresponding nonequilibrium occupation determined by $\Sigma^</\ii\Gamma = \tfrac{1}{2}(1 - \sigma_z)$ at zero temperature if one assumes fully occupied (empty) undriven valence (conduction) band. The upper and lower band nonequilibrium occupations are $f_{\pm} = \tfrac{1}{2} \left(1 \mp \cos \theta \right)$, and this shows that the upper band occupation increases upon increasing $\Omega$, i.e. the band overlap $\delta = - d_z$, whereas the lower band occupation decreases. This can be intuitively understood as a transfer of electrons from the lower to upper band as the optical pumping increases the overlap between the two bands. The Berry curvature of the two Floquet bands reads
\begin{eqnarray*}
\label{Berry curv}
\vec{F}_{+}(\vec{k}) = \cos^2 \frac{\theta}{2} \nabla \times \vec{a}_1 + \sin^2 \frac{\theta}{2} \nabla \times \vec{a}_2 - \cos^2 \frac{\theta}{2} \nabla \times \nabla \phi \\
\vec{F}_{-}(\vec{k}) = \sin^2 \frac{\theta}{2} \nabla \times \vec{a}_1 + \cos^2 \frac{\theta}{2} \nabla \times \vec{a}_2 - \sin^2 \frac{\theta}{2} \nabla \times \nabla \phi
\end{eqnarray*}
where $\vec{a}_1(\vec{k})$ and $\vec{a}_2(\vec{k})$ are the Berry connections for the original bands in the absence of optical pumping. Hence, the additional contribution of the photoinduced Weyl points to the Hall conductivity is $\sigma_{yz}^{(p)} = \sigma_{yz} - \sigma_{yz}^{0}$, with $\sigma_{yz}^{0}$ being the Hall conductivity in the absence of optical pumping, which is obtained by setting $\theta = 0$.
\begin{equation*}
\label{Photo sigmayz}
\sigma_{yz}^{(p)}  =  \frac{e^2}{2\pi h} \sum_{\vec{k} \in \mathrm{2D \; BZ}} \frac{1}{4}\sin^2 \theta  \nabla \times \left( \nabla \phi + \vec{a}_2 - \vec{a}_1  \right).
\end{equation*}
The effect of $\Gamma$ is included explicitly in the occupation factor $\sin^2 \theta = \tfrac{V^2 |v^x|^2}{d_z^2 + V^2 |v^x|^2 + \Gamma^2}$. Therefore, at zero temperature, for $\Gamma \ll V |v^x|$, we obtain $\sigma_{yz}^{(p)} \sim \pi V |v^x| \delta(d_z) \propto E$, and for $\Gamma \gg V |v^x|$, $\sigma_{yz}^{(p)} \sim \tfrac{V^2 |v^x|^2}{\Gamma^2} \propto E^2$.

The Berry-curvature approach we used in Sec. \ref{Sec_conductivity} is in fact a bypass avoiding a complicated solution of the Keldysh Green's functions with damping rate $\Gamma$ present, i.e., $\Gamma$ is introduced to reflect the thermal coupling but does not enter the final result, which corresponds to the situation when $\Gamma \ll V$. And Hall conductivity should be proportional to $E$ as discussed above. On the other hand, within the same calculation framework, if we formally add a concrete damping $\ii\Gamma/2$ as the imaginary part of self-energy into the squared energy denominator of Berry curvature Eq.~\eqref{eq:BerryCurvature}, we are able to recover the desired $\sigma\propto E^2$ dependence, corresponding to the situation when $\Gamma$ is large enough compared with $V$. 

\begin{figure}  
\centering
\subfloat[$\sigma\propto V$]{
\label{fig:Vlinear} 
\begin{minipage}[c]{0.45\textwidth}
  \scalebox{0.5}{\includegraphics{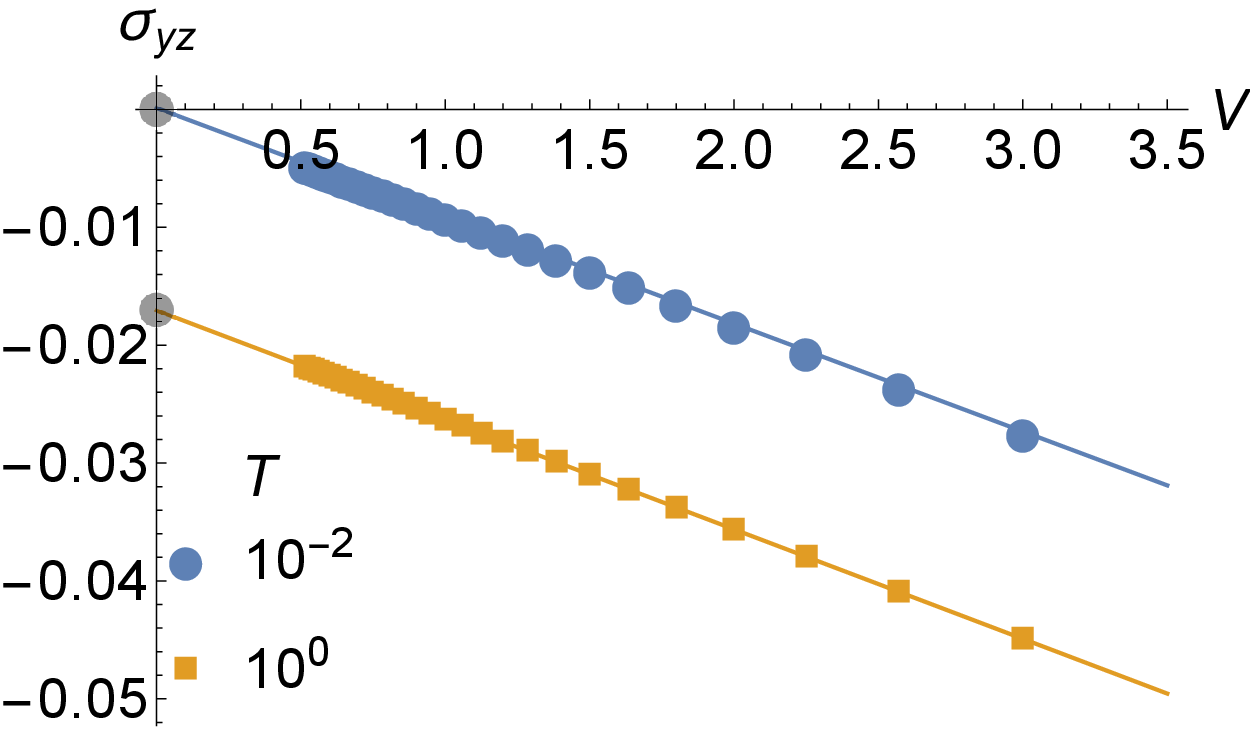}}
\end{minipage}}\hfill
\subfloat[$\sigma\propto V^2$]{
\label{fig:Vquadratic} 
\begin{minipage}[c]{0.45\textwidth}
  \scalebox{0.5}{\includegraphics{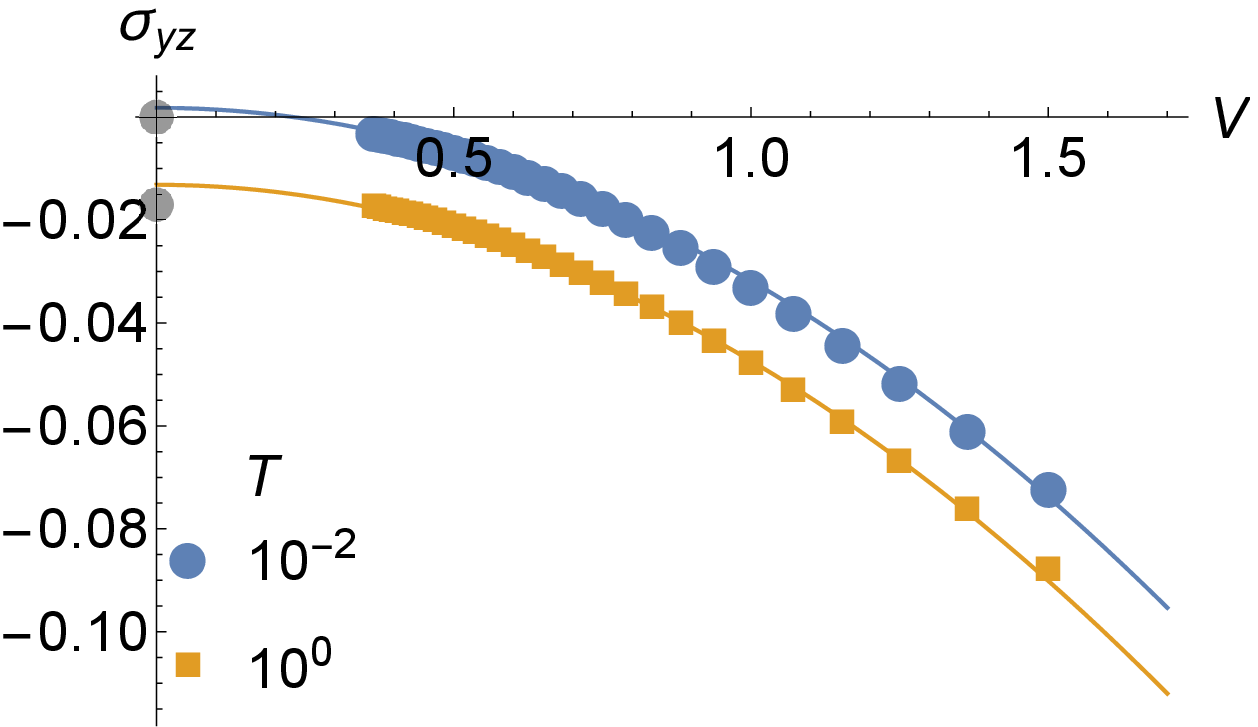}}
\end{minipage}}\hfill
\caption{(Color online) Hall conductivity's dependence on the strength of electric field $V$. Translucent grey dots are equilibrium conductivities without light irradiation. Other dots are numerically calculated data while lines are obtained from curve fitting. Temperature $T$ is measured in units of hopping energy $t_x$ in the Hamiltonian.}\label{fig:HallV}
\end{figure}

Towards this end, as shown in Fig.~\ref{fig:HallV}, we calculate the Hall conductivity in these two ways at 30 different driving strength $V$'s with a fixed driving frequency $\Omega=15.0$ (the results hold in general with no $\Omega$ dependence) and fit them with the method of least squares using a linear function and a quadratic function, respectively. In Fig.~\ref{fig:Vlinear} (Fig.~\ref{fig:Vquadratic}), $V$ is in the range $[\frac{\Omega}{30},\frac{\Omega}{5}]$ ($[\frac{\Omega}{42},\frac{\Omega}{10}]$) and we get a linear (quadratic) behavior where the coefficient of determination is almost unity for both of the two typical low and high temperatures. In addition to such a confirmation of the theoretical expectation, it is also helpful to point out several other properties. First of all, the conductivities do not reverse sign if we change the sign of the driving electric field $E$ or $V$. Secondly, the fitted slope does not have a considerable temperature dependence simply because the temperature-dependent occupation function will not be affected by the driving strength. However, the absolute value of the conductivity intercept indeed increases with temperature, reflecting the residual conductivity caused by the nonequilibrium thermal occupation. Also the intercepts approach to the equilibrium conductivities better in the linear case simply because our quadratic calculation is in fact an approximation and only for the sake of formally recovering the quadratic dependence. Lastly, in the latter case, $V$ is taken to be small enough compared with the damping $\Gamma=\frac{\Omega}{3}$ we used. Other calculation at different $\Gamma$'s shows that the absolute value of conductivity decreases with increasing $\Gamma$. This asserts that the longer the relaxation time is, the larger nonlinear effect arises from photo-excited electrons.

\section{Concluding remark}
Let us briefly discuss the experimental relevance of our work by providing some realistic estimates. Typical semiconductor materials and Weyl semimetals\cite{seeWeyl1} can have a band gap of about $0.5\mathrm{eV}$; thus, the frequency of the optical pumping needed to close this gap is $\Omega\sim0.75\times10^3\mathrm{THz}$. To generate a first-order dynamically opened gap at anticrossings of about $5\mathrm{meV}$, one needs a driving energy of $V\sim0.05\mathrm{eV}$. For a typical electric dipole moment of $1\mathrm{Debye}$, this can be achieved with an electric field strength $E\sim0.02\mathrm{V\cdot\AA^{-1}}$. To distinguish this gap and pick up the Berry phase correctly, this also means the ambient temperature should be less than $50\mathrm{K}$. With the typical value $0.5\frac{e^2}{a_0h}$ of Hall conductivity calculated in this work, and an assumed lattice constant $a_0=4\mathrm{\AA}$, this gives a $\sigma_{yz}\sim 1\times10^5(\Omega\cdot m)^{-1}$. We thus conclude that the photoinduced Weyl points should be detectable using ultrafast pump-probe angle-resolved photoemission spectroscopy\cite{Gedik,Gedik2,detect3}, and the anomalous Hall conductivity is also within the reach of current transport measurement for samples of mesoscopic scale.

In summary, we show the possibility of generating Weyl points by applying linearly polarized light, which allows us to optically control the anomalous Hall conductivity of the system. Using a tight-binding model, it is explicitly shown that, starting either from a band insulator or a Weyl semimetal, such an irradiation can generate new photoinduced Weyl points by optically bridging conduction and valence bands. In this model, these Weyl points occur along three different high symmetry lines, whose classification and topological charge explain the numerically and analytically obtained Chern number phase diagrams. The anomalous Hall conductivity is calculated via a Floquet-Keldysh formalism, which allows us to include the crucial effects of nonequilibrium occupation of the different Floquet bands. The anomalous Hall conductivity shows a nonmonotonous variation with regards to the optical driving frequency, due to the creation/annihilation and momentum-space shift of photoinduced monopoles and antimonopoles. The conductivity dependence on driving strength is also studied, which complies with the general scaling rules of nonlinear optical effects. These nonlinear driven effects are expected to be experimentally detectable, especially in anomalous Hall effect measurements.
\\ 
\\ 
\\ 
\textit{\textbf{Note added}}   -
Recently, we noticed several different works posted on arXiv or just published. Dissimilarly, they use circularly polarized light to create Weyl points from a single Dirac point\cite{Rubio} or from a nodal line semimetal\cite{ZhongWang,Awadhesh}. A 2D version of driven Weyl semimetal phase was also considered in arrays of weakly coupled wires\cite{DanielLoss}.

\section*{Acknowledgments}
X.-X.Z thanks Ming Lu and Takahiro Morimoto for helpful discussions. X.-X.Z was partially supported by the ALPS program and by the Grant-in-Aid for JSPS Fellows (No. 16J07545). This work was supported by JSPS Grant-in-Aid for Scientific Research Grant (No. 24224009) and by JSPS Grant-in-Aid for Scientific Research on Innovative Areas Grant (No. 26103006). This work was also supported by CREST, Japan Science and Technology Agency.

\label{Bibliography}
\bibliography{reference.bib}  

\end{document}